\documentclass[11pt]{article}
\pdfoutput=1
\usepackage{amsmath,amssymb,graphicx}
\usepackage{hyperref}
\usepackage[T1]{fontenc}
\usepackage[charter]{mathdesign}
\usepackage[font=small,labelfont=bf]{caption}
\usepackage{cite}

\newcommand{\beq}{\begin{eqnarray}}
\newcommand{\eeq}{\end{eqnarray}}
\def\simlt{\stackrel{<}{{}_\sim}}
\def\simgt{\stackrel{>}{{}_\sim}}

\usepackage[usenames,dvipsnames]{xcolor}

\title{\bf Vacuum Instabilities with a Wrong-Sign Higgs-Gluon-Gluon Amplitude}
\author{Matthew Reece \\
{\small \texttt{mreece@physics.harvard.edu}}\\
\em{Department of Physics, Harvard University, Cambridge, MA 02138}}

\begin{document}
\maketitle

\begin{abstract}
The recently discovered 125 GeV boson appears very similar to a Standard Model Higgs, but with data favoring an enhanced $h \to \gamma \gamma$ rate. A number of groups have found that fits would allow (or, less so after the latest updates, prefer) that the $ht{\bar t}$ coupling have the opposite sign. This can be given meaning in the context of an electroweak chiral Lagrangian, but it might also be interpreted to mean that a new colored and charged particle runs in loops and produces the opposite-sign $hGG$ amplitude to that generated by integrating out the top, as well as a contribution reinforcing the $W$-loop contribution to $hFF$. In order to not suppress the rate of $h \to WW$ and $h \to ZZ$, which appear to be approximately Standard Model-like, one would need the loop to ``overshoot,'' not only canceling the top contribution but producing an opposite-sign $hGG$ vertex of about the same magnitude as that in the SM. We argue that most such explanations have severe problems with fine-tuning and, more importantly, vacuum stability. In particular, the case of stop loops producing an opposite-sign $hGG$ vertex of the same size as the Standard Model one is ruled out by a combination of vacuum decay bounds and LEP constraints. We also show that scenarios with a sign flip from loops of color octet charged scalars or new fermionic states are highly constrained.
\end{abstract}

\section{Introduction}
\label{sec:intro}

The Higgs discovery represents a major milestone in particle physics~\cite{ATLAS:2012gk,CMS:2012gu}. It brings renewed urgency to the question of naturalness: if the Higgs has precisely the properties predicted by the Standard Model, we may be forced to confront the possibility that we live in what is, to all appearances, a finely-tuned world. The experimental results so far present us with tantalizing hints that $\sigma \times~{\rm Br}(h \to \gamma \gamma)$ may be substantially larger than the Standard Model prediction~\cite{CMS-PAS-HIG-12-015,ATLAS-CONF-2012-091}. Indeed, a number of groups of theorists have attempted to fit the data allowing for non-Standard-Model Higgs couplings, both before~\cite{Englert:2011aa,Carmi:2012yp,Azatov:2012bz,Espinosa:2012ir,Giardino:2012ww,Li:2012ku,Rauch:2012wa,Ellis:2012rx,Klute:2012pu,Azatov:2012wq,Carmi:2012zd} and after~\cite{Corbett:2012dm,Giardino:2012dp,Buckley:2012em,Ellis:2012hz,Montull:2012ik,Espinosa:2012im,Carmi:2012in,Plehn:2012iz,Espinosa:2012in} the July 4, 2012 discovery announcement. 

Although many details of the fits and the allowed parameter space are explained in these references, we can summarize the situation (keeping in mind that the error bars are still rather large) by saying that the Higgs $\sigma \times~{\rm Br}$ to $WW$ and $ZZ$ is essentially consistent with the Standard Model, the rate to $\gamma\gamma$ is somewhat high, and the rate to $\tau$ leptons may be low although Tevatron results suggest that the $b$-quark rate is not very suppressed. In almost every way, the Higgs appears to be nearly Standard-Model-like. Nonetheless, fits of the Higgs couplings allow (or even, less so after recent ATLAS $h \to WW$ results, favor) a region with $R_t = -1$, i.e. a flipped sign of the Higgs--top--top coupling. This sign is fixed in the Standard Model without higher-dimension operators, but can be altered in the electroweak chiral Lagrangian. Another interpretation, however, could be that new particles run in the loop for both $h \to gg$ and $h \to \gamma\gamma$ with the opposite sign of the top.

\begin{figure}[h]
\begin{center}
\includegraphics[width=0.9\textwidth]{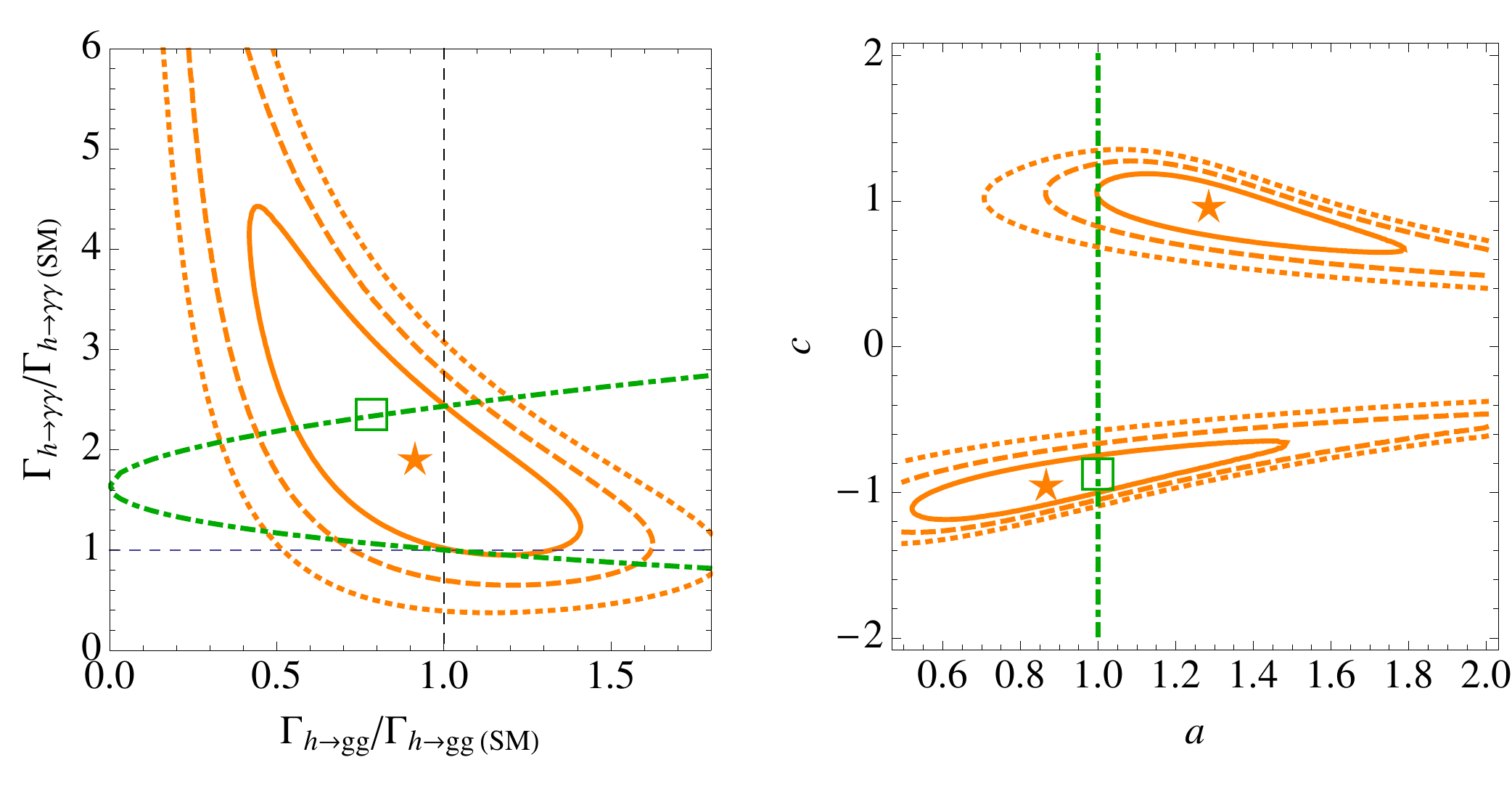}
\end{center}
\caption{Fit of the $WW$, $ZZ$, and $\gamma\gamma$ channels in the ATLAS and CMS 7+8 TeV data, allowing the $hGG$ and $hFF$ amplitudes to vary. The best-fit point is marked with the orange star, which is surrounded by 1, 2, and 3$\sigma$ orange contours. The Standard Model value (1,1) falls on the $1\sigma$ contour. The green dot-dashed curve illustrates the possible contributions from top partners, with the best-fit point along this curve marked with the open green square. The left- and right-hand plots show the same information, at left in the plane of $gg$ and $\gamma\gamma$ partial widths and at right in the plane of coefficients $a$ and $c$ for Higgs couplings to vectors and fermions, respectively. This fit is for illustrative purposes only; the reader can find fits incorporating more channels and more thorough statistical treatments in the literature.} 
\label{fig:higgsfitplot}
\end{figure}

To illustrate the possibility of achieving a better fit to the data with new colored and charged particles, we have performed a simple fit to the CMS and ATLAS combined 7 and 8 TeV data in the $\gamma\gamma$~\cite{CMS-PAS-HIG-12-015,ATLAS-CONF-2012-091}, $ZZ$~\cite{ATLAS-CONF-2012-092,CMS-PAS-HIG-12-016}, and $WW$~\cite{CMS-PAS-HIG-12-017,ATLAS-CONF-2012-098} channels, shown in Figure~\ref{fig:higgsfitplot}. Our fit uses six experimental inputs with two parameters, so we use $\Delta \chi^2_{4~{\rm d.o.f.}} = 4.72, 9.72,$ and $16.25$ to define the $1\sigma$, $2\sigma$, and $3\sigma$ contours. Because our goal is to illustrate a qualitative point more than to extract precision information from the data, we omit other decay modes as well as vector boson fusion and other production channels. Furthermore, we do not take all signal strength values at the same mass, for instance taking the ATLAS $\gamma\gamma$ channel $\sigma \times {\rm Br}$ to be $1.9 \pm 0.5$ times the Standard Model rate despite the fact that this value is attained for a signal hypothesis of $m_h = 126.5$ GeV whereas other channels we take into account have $m_h = 125$ GeV. Nonetheless, this simple fit gives a similar result to the many other recent analyses, with the best-fit point having a slightly smaller $hGG$ coupling $\Gamma(h \to gg) = 0.9~\Gamma_{\rm SM}(h \to gg)$ and a substantially larger $h\gamma\gamma$ coupling $\Gamma(h \to \gamma \gamma) = 1.9~\Gamma_{\rm SM}(h \to \gamma \gamma)$. Note that the recent ATLAS $WW$ result~\cite{ATLAS-CONF-2012-098}, with an observed rate (combining 7 and 8 TeV data) of $1.4 \pm 0.5$ times the Standard Model expectation, partially counteracts the tendency of previous $h \to WW$ searches to prefer a diminished gluon fusion rate. 

Figure~\ref{fig:higgsfitplot} also shows a curve of values that can be obtained with stops running in loops. Beginning from the SM point $(1,1)$ and moving to the left, one sees that initially increasing the $h \to \gamma\gamma$ rate decreases the $h \to gg$ rate, but at a certain point the curve turns around and both rates increase. This corresponds to reversing the sign of the $hGG$ amplitude. The best-fit point on the curve has $\Gamma(h \to gg) = 0.8~\Gamma_{\rm SM}(h \to gg)$ (but with an amplitude of opposite sign) and $\Gamma(h \to \gamma\gamma) = 2.3~\Gamma_{\rm SM}(h \to \gamma\gamma)$. A similar observation appeared recently in Ref.~\cite{Buckley:2012em}. However, it is important to realize that this is a very large loop effect, inverting the sign of the $hGG$ amplitude from a top loop by subtracting a new contribution twice as large. Such large loop effects are not expected in the ``natural SUSY'' scenario that often motivates consideration of light stops~\cite{Dimopoulos:1995mi,Cohen:1996vb}, and are not innocuous. In particular, the same particles that run in these loops affect the Higgs potential, and if they are scalars, they have a potential of their own with possible new minima. We will argue that these effects are not benign, and trying to use the upper branch of the green curve in Figure~\ref{fig:higgsfitplot} to explain the data brings with it a host of new problems, whether the new particles are scalars or fermions.

\section{Loop effects of charged and colored particles}
\label{sec:loops}

\subsection{Computing the effects}
\label{sec:computing}

New particles that obtain a portion of their mass from the Higgs boson also alter the Higgs potential. We will be primarily concerned with their effect on the Higgs quartic, which determines the mass of the Higgs boson once the appropriate vacuum is found. To compute the shift in the quartic, we use the one-loop Coleman-Weinberg potential,
\beq
V_{\rm CW} = \frac{1}{32\pi^2} \sum (-1)^F {\cal M}^4 \left(\log\frac{{\cal M}^2}{\mu^2} - \frac{3}{2} \right),
\eeq
expressing the mass of the new particles in terms of the Higgs field $H$, expanding as a function of $H$, and reading off the coefficient of $\left|H\right|^4$ to obtain a correction $\delta \lambda$ to the quartic. Note that when expanding around the origin and reading off the $\left|H\right|^4$ term, we neglect possible $\left|H\right|^4 \log\left|H\right|^2$ terms that would arise from fields that are massless when the Higgs has no vev. Because we are interested in negative contributions to the $hGG$ coupling, the dominant effect of increasing the Higgs vev should be to {\em decrease} the mass of the fields we integrate out, and this is a reasonable approximation to use.

Corrections to the effective Higgs couplings to photons and gluons are easily understood in terms of the low-energy theorem~\cite{Ellis:1975ap,Shifman:1979eb}. Namely, to read off the effective coupling induced by integrating out heavy particles, one treats them as a Higgs-dependent mass threshold in the beta function, obtaining the effective vertex from the running of $1/g^2$:
\beq
-\frac{1}{4 g^2} G^a_{\mu \nu} G^{a \mu\nu} \supset -\frac{1}{4} \left(-\frac{\Delta b}{16\pi^2} \log \det M^2(h) \right) G^a_{\mu\nu} G^{a\mu\nu} \supset \frac{\Delta b}{64\pi^2} \frac{h}{v} G^a_{\mu\nu}G^{a\mu\nu} \frac{\partial \log \det M^2}{\partial \log v}.
\eeq
with $\Delta b$ the beta function coefficient of the states that were integrated out. An analogous statement holds for couplings to photons, the only difference being that it is the electromagnetic beta function coefficient that appears. In the case that the mass of the new particles is not much greater than half the Higgs mass, it can be important to take into account mass-dependent corrections to the low-energy theorem. In particular, for fermions these corrections are $1 + \frac{7 m_h^2}{120 m_F^2} + {\cal O}(m_h^4/m_F^4)$ and for scalars $1 + \frac{2 m_h^2}{15 m_S^2} + {\cal O}(m_h^4/m_S^4)$.

\begin{figure}[h]
\begin{center}
\includegraphics[width=0.46\textwidth]{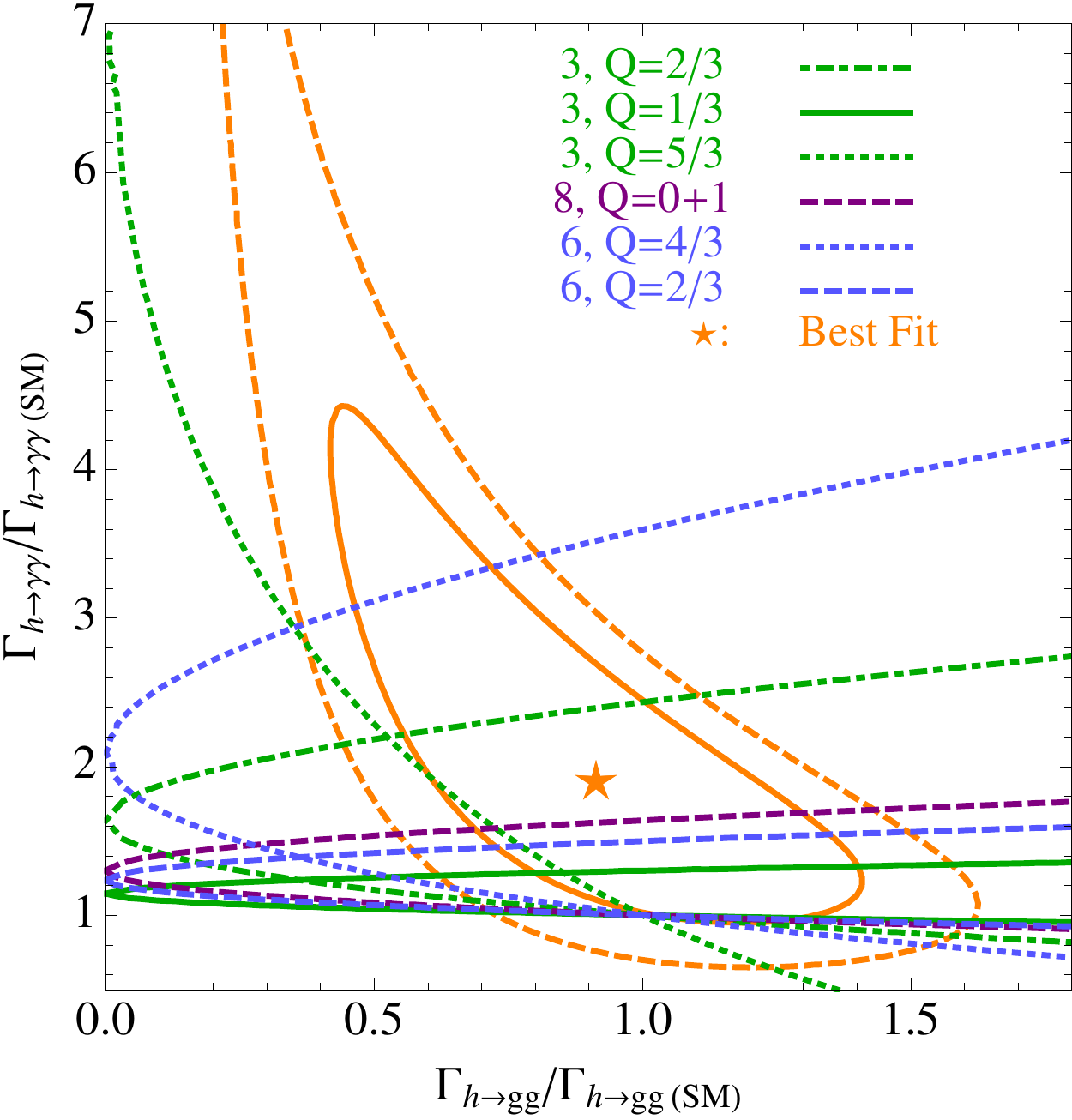}
\end{center}
\caption{Fit of the $WW$, $ZZ$, and $\gamma\gamma$ channels in the ATLAS and CMS 7+8 TeV data, with $1\sigma$ and $2\sigma$ contours as in the left-hand plot of Figure~\ref{fig:higgsfitplot}, but now showing the values achieved by adding particles in the loop in a variety of representations of SU(3)$_c$ and U(1)$_{\rm EM}$.} 
\label{fig:higgsoptions}
\end{figure}

If we have a new colored state that carries charge $Q$ and is in an SU(3)$_c$ representation with quadratic Casimir $C_2(R)$, we can evaluate its effect by rescaling the contribution of the top loop amplitude in the Standard Model. Namely, defining
\beq
R_g = \frac{\Gamma(h \to gg)}{\Gamma_{\rm SM}(h \to gg)}~~{\rm and}~~R_\gamma = \frac{\Gamma(h \to \gamma\gamma)}{\Gamma_{\rm SM}(h \to \gamma\gamma)},
\eeq
we find that they are related by:
\beq
R_\gamma = \left[1 + 0.28 \xi \left(1 \mp \sqrt{R_g}\right)\right]^2,
\eeq
where the sign of the square root is determined by the sign of the $hGG$ amplitude, and
\beq
\xi = \frac{Q^2}{C_2(R)} \frac{C_2({\bf 3})}{Q_{\rm top}^2} = \frac{3 Q^2}{C_2(R)}.
\eeq
As discussed recently in, for instance, Ref.~\cite{Cohen:2012wg}, the choices of charge and representation are fairly restricted by needing particles that can decay to the Standard Model (given the lack of detected stable particles of exotic charge). We plot the possible effects of several examples of plausible charge assignments in Figure~\ref{fig:higgsoptions}. Each of the curves has two branches meeting at $R_g = 0$, with the upper branch corresponding to the case with an inverted sign for the $hGG$ amplitude. Notice that charge-$2/3$ color triplets can improve the fit, but other charges for color triplets are of little help in the inverted sign regime. (The charge $5/3$ triplet discussed in Ref.~\cite{Cohen:2012wg} may help the fit slightly, but with the sign of $hGG$ not altered and hence the overall Higgs production rate decreased. This predicts that the measured rate for $h\to WW, ZZ$ should decrease in the future.) The combination of a neutral and charge $1$ color octet with the same mass can give an interesting improvement in the fit. A color sextet of charge $2/3$ can also offer some improvement. (As this paper was nearing completion, Ref.~\cite{Dorsner:2012new} appeared advocating color octet or sextet scalars with opposite-sign $hGG$ amplitude as a hint of unification. Given the vacuum stability and tuning arguments discussed below, we are much less sanguine.)

A note on conventions: we will follow~\cite{Martin:1997ns} in taking $v \approx 174$ GeV. Our choices are such that $y_t \approx 1$, $m_W^2 = \frac{1}{2} g^2 v^2$, and $m_Z^2 = \frac{1}{2}\left(g^2 + g'^2\right)v^2$.

\subsection{New fermionic states}
\label{sec:fermions}

Let us first consider the case of new fermionic states. We assume two vectorlike pairs of fermions, $\psi, {\bar \psi}$ and $\chi, {\bar \chi}$ with charges such that Yukawa couplings $H {\bar \psi} \chi$ and $H^\dagger {\bar \chi} \psi$, so that the mass matrix in the basis $\psi,\chi,{\bar \psi},{\bar \chi}$ is:
\beq
{\cal M}_F & = &  \left(\begin{matrix} 0 & M_F^T \\ M_F & 0 \end{matrix}\right), \\
{\rm where:}~~M_F & = & \left(\begin{matrix} m_\psi & y_1 v \\ y_2 v & m_\chi \end{matrix}\right).
\eeq
In this case, the correction to the $h \to gg$ amplitude, relative to the Standard Model amplitude from a top loop (and neglecting mass effects) is:
\beq
\frac{\delta A(hGG)}{A_{\rm SM}(hGG)} = 2 \frac{\Delta b_{\bf r}}{\Delta b_{\bf 3}} \left(1 - \frac{m_\chi m_\psi}{m_\chi m_\psi - y_1 y_2 v^2}\right).
\eeq
In particular, because there are vectorlike masses that are split by the mixing terms proportional to the Yukawas, we get a negative contribution to the amplitude. The factor $\frac{\Delta b_{\bf r}}{\Delta b_{\bf 3}}$ is the ratio of the SU(3)$_c$ beta-function coefficient of the representation that $\psi$ and $\chi$ transform under, relative to the beta-function coefficient of a triplet. 

Loops of fermions contribute a correction to the Higgs quartic, which in the special case $m_\psi = m_\chi = m$ is:
\beq
\delta \lambda_F = -\frac{N_{c;F}}{16\pi^2} \left\{\left(y_1^4 + y_2^4\right) \log \frac{m^2}{\mu^2} + \frac{1}{6}\left(5y_1^2 -2 y_1 y_2 + 5y_2^2\right)\left(y_1+y_2\right)^2 \right\}.
\eeq
The result in the more general case $m_\chi \neq m_\psi$ is listed in Appendix~\ref{app:CW}. Notice that the logarithmic term here can be interpreted as encoding a beta function coefficient. Because the full renormalized potential must be independent of $\mu$, the tree-level quartic must run in such a way as to cancel the $\mu$-dependence of the Coleman-Weinberg potential. 

\subsection{New scalar states}
\label{sec:scalars}

Assume a mass matrix
\beq
{\cal M}_S^2 = \left(\begin{matrix} m_1^2 + \lambda_1 v^2 & A v \\ A v &  m_2^2 + \lambda_2 v^2 \end{matrix}\right).
\eeq
The correction to the $h \to gg$ amplitude relative to the Standard Model amplitude is
\beq
\frac{\delta A(hGG)}{A_{\rm SM}(hGG)} = \frac{\Delta b_{\bf r}}{\Delta b_{\bf 3}} \frac{v^2 \left(\lambda_1 m_2^2 + \lambda_2 m_1^2 + 2 \lambda_1 \lambda_2 v^2 - A^2\right)}{4\left(\left(m_1^2 + \lambda_1 v^2\right)\left(m_2^2 + \lambda_2 v^2\right) - A^2 v^2\right)}, \label{eq:hGGscalarcorrection}
\eeq
The factor of $1/4$ arises from the relative beta function coefficients of a single color-triplet scalar and the top quark, whereas the factor $\frac{\Delta b_{\bf r}}{\Delta b_{\bf 3}}$ again corrects for the case when the field is not in the ${\bf 3}$ of SU(3)$_c$. As in the case of fermions, the effect of mixing (here proportional to $A$) is to split the mass eigenstates and thus give a negative contribution. On the other hand, the quartic couplings $\lambda_{1,2}$ can give contributions of either sign. 

The correction to the Higgs quartic in the case where the mass parameters $m_1$ and $m_2$ are equal is:
\beq
\delta \lambda_S = \frac{N_{c;S}}{32\pi^2} \left((\lambda_1^2 + \lambda_2^2)\log\frac{m^2}{\mu^2} + (\lambda_1 + \lambda_2) \frac{A^2}{m^2} - \frac{1}{6} \frac{A^4}{m^4}\right).
\eeq
The result in the more general case $m_1 \neq m_2$ is given in Appendix~\ref{app:CW}. To compare to a more familiar expression: if the scalar states are stops in a supersymmetric theory, we have $m_1^2 = m_{Q_3}^2$, $m_2^2 = m_{u^c_3}^2$, $N_{c;S} = 3$, $\lambda_1 = y_t^2 + \left(\frac{1}{2}-\frac{2}{3}\sin^2\theta_W\right) \cos(2\beta) \frac{g^2 + g'^2}{2}$, $\lambda_2 = y_t^2 + \frac{2}{3}\sin^2 \theta_W \cos(2\beta) \frac{g^2 + g'^2}{2}$, and $A = y_t \left(A_t \sin \beta - \mu \cos \beta\right)$. In particular, the part of $\delta \lambda_S$ that is polynomial in $A$, dropping terms of order $g^2$, taking $m_{Q_3}^2 = m_{u^c_3}^2 = m_{\tilde t}^2$, and assuming large enough $\tan \beta$, is:
\beq
\delta \lambda_S \approx \frac{3}{16\pi^2} y_t^4 \left(\frac{X_t^2}{m_{\tilde t}^2} - \frac{1}{12}  \frac{X_t^4}{m_{\tilde t}^4}\right),
\eeq
with $X_t = A_t - \mu \cot \beta$. This is the familiar result that can be found in, for example,~\cite{Carena:1995wu}. As for the logarithmic term, tops contribute $-\frac{3}{16\pi^2}y_t^4 \log \frac{m_t^2}{\mu^2}$, so the $\mu$-dependence cancels and the leftover logarithmic correction is $\frac{3}{16\pi^2} y_t^4 \log\frac{m_{\tilde t}^2}{m_t^2}$, which is also of the familiar expected form.

\section{Vacuum stability}
\label{sec:vacuum}

Given the results of the Coleman-Weinberg calculation, it is apparent that trying to achieve a large enough loop correction to change the sign of the $hGG$ coupling is a dangerous game. Flipping the sign implies having a particle with a mass that diminishes with increasing Higgs VEV. One possibility for this is a mixing effect: either one has vectorlike fermions getting a majority of their mass independent of the Higgs, or scalars that mix analogously to the familiar case of stops in supersymmetric theories. In the case of fermions, the most dangerous effect is the renormalization group running from the fermion Yukawa coupling, which pushes the Higgs quartic toward negative values in the UV and can lead to an unstable vacuum~\cite{ArkaniHamed:2012kq}. For scalars, the RG effect is not dangerous, as the Higgs quartic is pushed toward larger values in the UV. However, there is a large negative threshold correction, proportional to the fourth power of the mixing parameter $A$ (familiar from the case of stops), which threatens to make the Higgs tachyonic. Furthermore, such large mixing parameters can lead to color and charge breaking minima of the tree-level potential~\cite{Kusenko:1996jn}. The remaining alternative, which does not require large mixings, is that one can have scalars with a positive mass$^2$ and a negative quartic coupling to the Higgs. Such a negative quartic coupling again can lead to color- and charge-breaking minima or runaway directions. Our goal in this section is to give some simple estimates of the parameter space leading to catastrophic vacuum instabilities and show that most attempts to achieve an $hGG$ coupling of approximately the Standard Model magnitude but opposite sign are ruled out by them.

\subsection{Inverting $hGG$ with Stops}

Given that we are looking for large changes to the Higgs potential that require light new colored and charged particles, it is reasonable to first consider whether stops can be responsible, since naturalness of electroweak symmetry breaking in supersymmetric theories favors light stops~\cite{Dimopoulos:1995mi,Cohen:1996vb}. In the case of stops, the general results discussed in the previous section imply a correction to the $hGG$ amplitude (specializing the general result Eq.~\ref{eq:hGGscalarcorrection}):
\beq
\frac{A(hGG)}{A_{\rm SM}(hGG)} = 1 + \frac{1}{4} \left( \frac{m_t^2}{m_{{\tilde t}_1}^2} +  \frac{m_t^2}{m_{{\tilde t}_2}^2} - \frac{m_t^2 X_t^2}{m_{{\tilde t}_1}^2 m_{{\tilde t}_2}^2}\right),
\label{eq:stopggH}
\eeq
up to small $D$-term corrections (taken into account in the plots below). Here $m_{{\tilde t}_1}$ and $m_{{\tilde t}_2}$ are mass eigenvalues, not Lagrangian parameters. The effect of stops on Higgs branching ratios has been discussed in several papers in the recent literature~\cite{Carmi:2012yp,Blum:2012ii,Giardino:2012dp,Buckley:2012em,Espinosa:2012in}, which reach a variety of conclusions. As emphasized by Ref.~\cite{Blum:2012ii}, light unmixed stops tend to increase the $hGG$ coupling and decrease the $h\gamma\gamma$ coupling, whereas highly mixed stops contribute large corrections to $m_{H_u}^2$ (thus requiring more tuning for EWSB) and lead to large corrections to $b \to s \gamma$ that must also be tuned away. The same considerations led Ref.~\cite{Espinosa:2012in} to focus on the ``funnel'' region in which the stop corrections to $hGG$ are small. On the other hand, Refs.~\cite{Carmi:2012in,Buckley:2012em} argued for light and highly mixed stops in the region with the inverted sign of $hGG$, which could improve the fit to data. 

\begin{figure}[h]
\begin{center}
\includegraphics[width=0.9\textwidth]{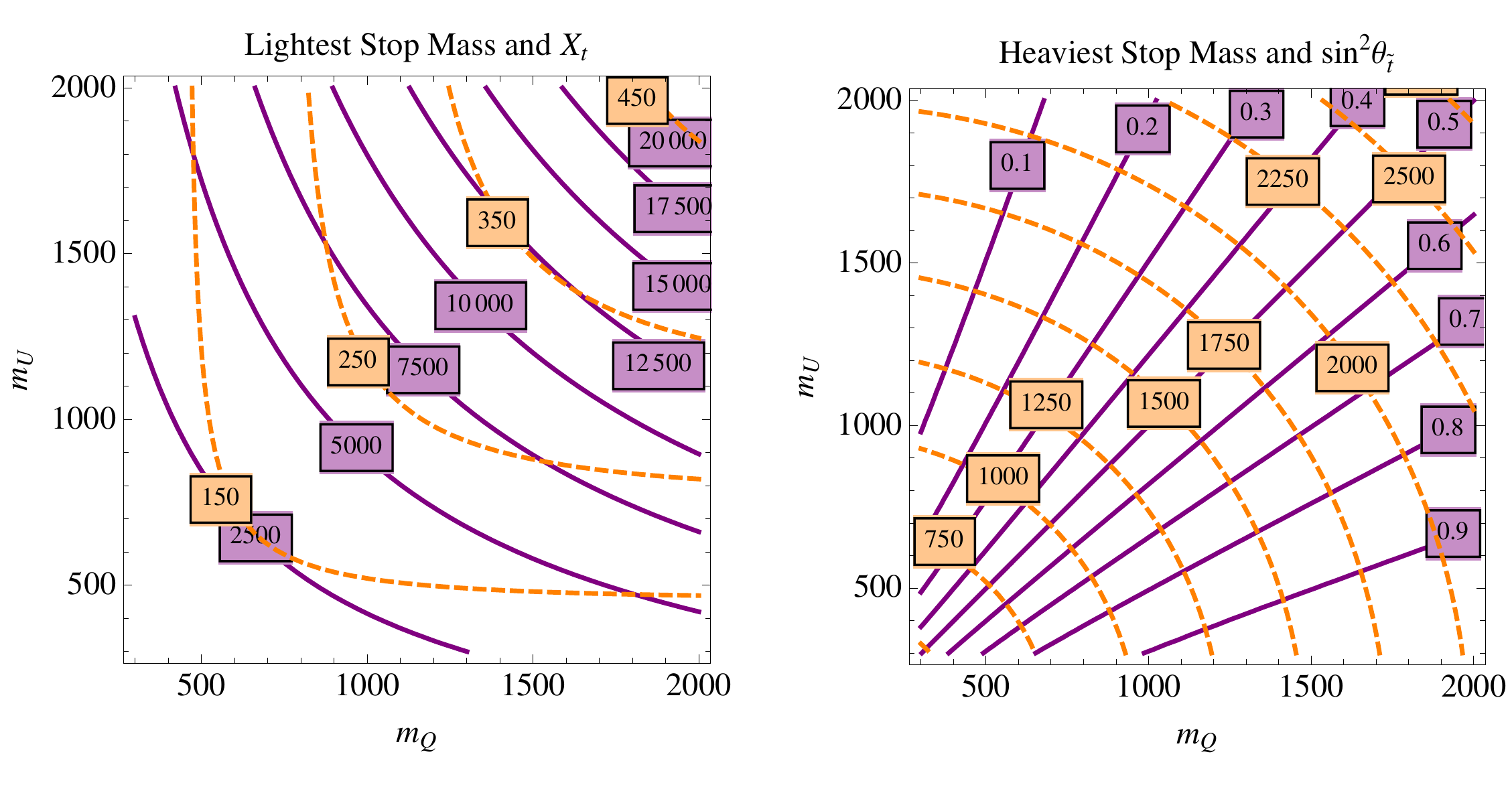}
\end{center}
\caption{Stop parameter space that achieves a $hGG$ coupling that is $-1$ times its Standard Model value. This condition reduces the three-dimensional parameter space $(m_Q, m_U, X_t)$ to two dimensions, which we parametrize with $m_Q$ and $m_U$. At left: contours of the lightest stop mass (orange, dashed) and the value of $X_t$ needed to achieve the desired coupling (purple, solid). At right: contours of the heavy stop mass (orange, dashed) and the corresponding stop mixing $\sin^2\theta_{\tilde t}$ parametrizing the right-handedness of the stop (purple, solid).} 
\label{fig:stopplane}
\end{figure}

We illustrate the parameter space that can achieve $A(hGG) = - A_{\rm SM}(hGG)$ in Figure~\ref{fig:stopplane}. As is clear from equation~\ref{eq:stopggH}, this occurs at very large values of the mixing parameter $X_t$. This leads to a large splitting between the two stop mass eigenstates. In this region of parameter space, the lightest stop eigenvalue tends to be fairly light. For example, pushing the light eigenstate up to 450 GeV implies 20 TeV $A$-terms, which is an enormously finely-tuned scenario, both from the point of view of electroweak symmetry breaking and of $b \to s \gamma$. In fact, from the Coleman-Weinberg discussion in Section~\ref{sec:scalars}, one can readily see that such large $A$-terms lead to very large negative threshold corrections to the Higgs mass. This implies the need for very large beyond-MSSM couplings of the Higgs boson that are capable of lifting its mass up to 125 GeV. When such couplings become large enough, it is difficult to imagine that other Higgs properties remain unmodified, so that considering only stop-loop modifications to the partial widths is dubious. On the other hand, one may wonder if the lower-left corner of the plot, with a light stop eigenstate, can fit the data, with large but no longer unreasonably large $A$-terms. It is still rather tuned. Recent experimental searches for direct production of light stops~\cite{ATLAS-CONF-2012-070,ATLAS-CONF-2012-071,ATLAS-CONF-2012-073,ATLAS-CONF-2012-074,CMS-PAS-SUS-12-009,CMS-PAS-SUS-11-022} constrain much of the stop parameter space with $m_{{\tilde t}_1} \simlt 500$ GeV, but only for sufficiently light neutralinos. The more squeezed regime will be probed by a combination of traditional missing-$E_T$ signatures~\cite{Meade:2006dw,Han:2008gy,Plehn:2010st,Asano:2010ut,Plehn:2011tf,Bai:2012gs,Plehn:2012pr,Alves:2012ft,Kaplan:2012gd,Chen:2012uw} and spin correlations~\cite{Han:2012fw}, and even the case of $R$-parity violation may be constrained soon~\cite{Brust:2012uf}. Nonetheless, for the moment, these considerations still allow as a logical possibility that light, highly mixed stops significantly alter the Higgs properties.

\begin{figure}[h]
\begin{center}
\includegraphics[width=0.9\textwidth]{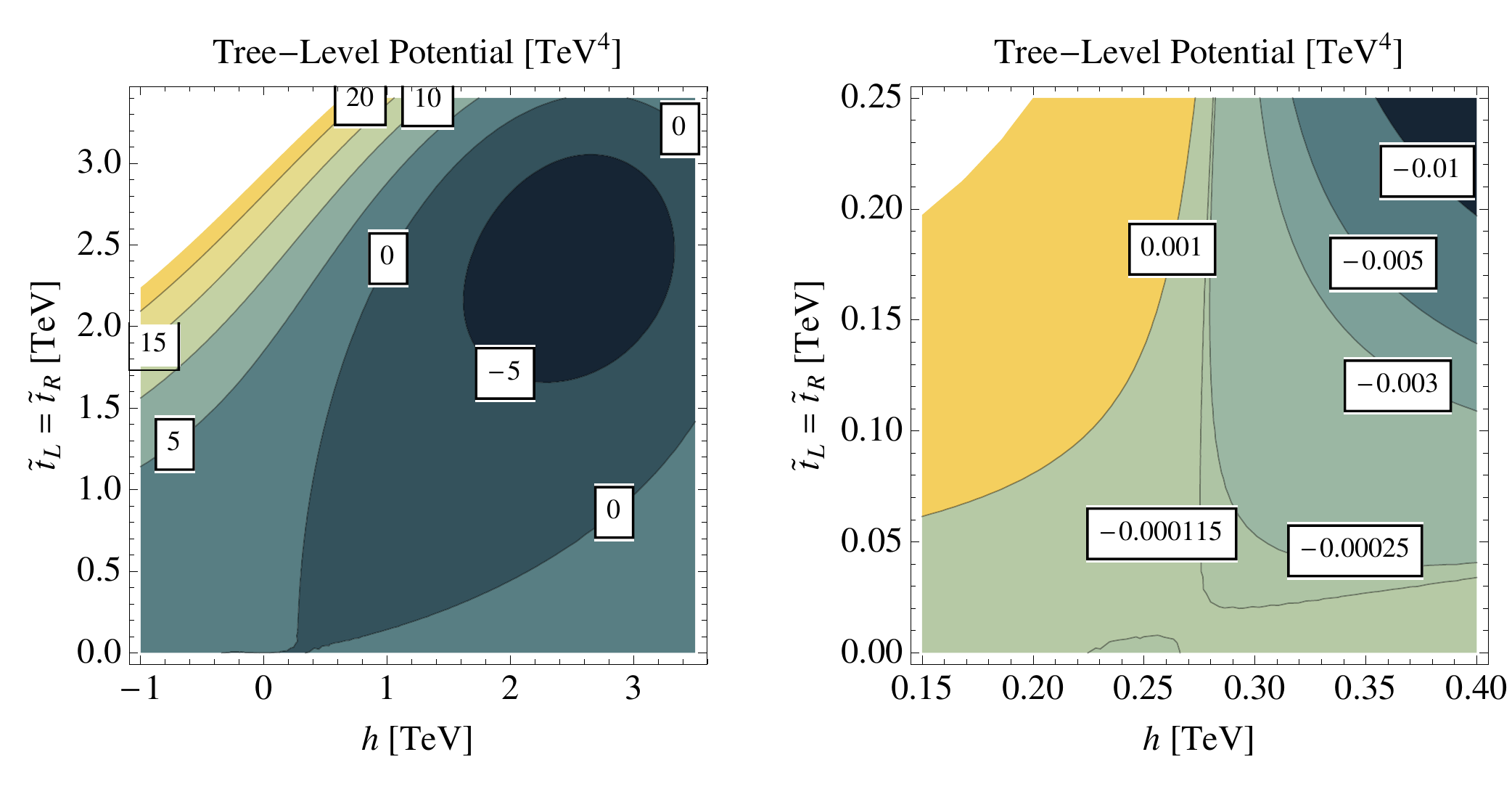}
\end{center}
\caption{Tree-level potential $V(h, {\tilde t}_L, {\tilde t}_R)$ along the subspace ${\tilde t}_L = {\tilde t}_R$. We have fixed $m_Q = m_U = 800$ GeV and adjusted $X_t$ to produce $A(hGG) = -A_{\rm SM}(hGG)$. The right-hand plot zooms in near the good EWSB vacuum where $\left<h\right> \approx 246$ GeV and the stops have no VEV. A much deeper minimum is located near the $D$-flat direction where the Higgs and stop VEVs are all equal. The barrier separating the two minima is shallow.} 
\label{fig:potentialcontours}
\end{figure}

However, vacuum instability poses an even more serious problem for this scenario than fine-tuning. The large $A$-term mixing is a trilinear scalar coupling ${\tilde t}_L {\tilde t}_R^* h$, so the potential can acquire large negative values when all three of these fields have VEVs. Because the Higgs and one stop eigenstate are relatively light, the barrier separating our EWSB vacuum from a color- and charge-breaking minimum can be relatively low. At large enough field values, quartic couplings arising from the Yukawa coupling will prevent the potential from being unbounded from below, even in the $D$-flat direction where the stop and Higgs VEVs are equal. Nonetheless, a deep charge- and color-breaking vacuum will exist when the $A$-term is large. This is illustrated with contour plots of the potential in Figure~\ref{fig:potentialcontours}. It remains to check whether the vacuum decay to this deep minimum happens fast enough to rule out this scenario.

For this calculation we use the tree-level potential for the up-type Higgs $H$ and the third generation squark superfields:
\beq
V(H, {\tilde Q}_3, {\tilde u}^c_3) & = & m_H^2 \left|H\right|^2 + m_Q^2 \left|{\tilde Q}_3\right|^2 + m_U^2 \left|{\tilde u}^{c}_3\right|^2 + y_t^2 \left(\left|{\tilde Q}_3 {\tilde u}^c_3\right|^2 + \left|H {\tilde Q}_3\right|^2 + \left|H {\tilde u}^c_3\right|^2\right) \nonumber \\
& + & \frac{1}{8} g'^2 \left(\left|H\right|^2 + \frac{1}{3} \left|{\tilde Q}_3\right|^2 - \frac{4}{3} \left|{\tilde u}^c_3\right|^2\right)^2 + \frac{1}{8} g^2 \left(\left|H\right|^2 - \left|{\tilde Q}_3\right|^2\right)^2 + \frac{4}{3} \left(\left|{\tilde Q}_3\right|^2 - \left|{\tilde u}^c_3\right|^2 \right)^2 \nonumber \\
& + & \delta \lambda \left|H\right|^4 - y_t X_t H {\tilde Q}_3 {\tilde u}^c_3 - \left(y_t X_t H {\tilde Q}_3 {\tilde u}^c_3\right)^*.
\eeq
We take $m_H^2 = -\frac{1}{2} m_h^2$ with $m_h = 125$ GeV the measured Higgs mass. Here $\delta \lambda$ represents the corrections required to achieve the appropriate measured Higgs VEV; we remain agnostic about what model generates these corrections (in particular, we do not tie them to the stop masses and the MSSM radiative corrections). In the plot in Figure~\ref{fig:potentialcontours}, we have taken the fields to be real valued, with $H = \frac{1}{\sqrt{2}} h, {\tilde Q}_3 = \frac{1}{\sqrt{2}} {\tilde t}_L$, and ${\tilde u}^c_3 = \frac{1}{\sqrt{2}} {\tilde t}_R$. We ignore the down-type Higgs; at large $\tan\beta$, it should not be important, and more generally we don't expect that it will qualitatively alter the results.

\begin{figure}[h]
\begin{center}
\includegraphics[width=0.9\textwidth]{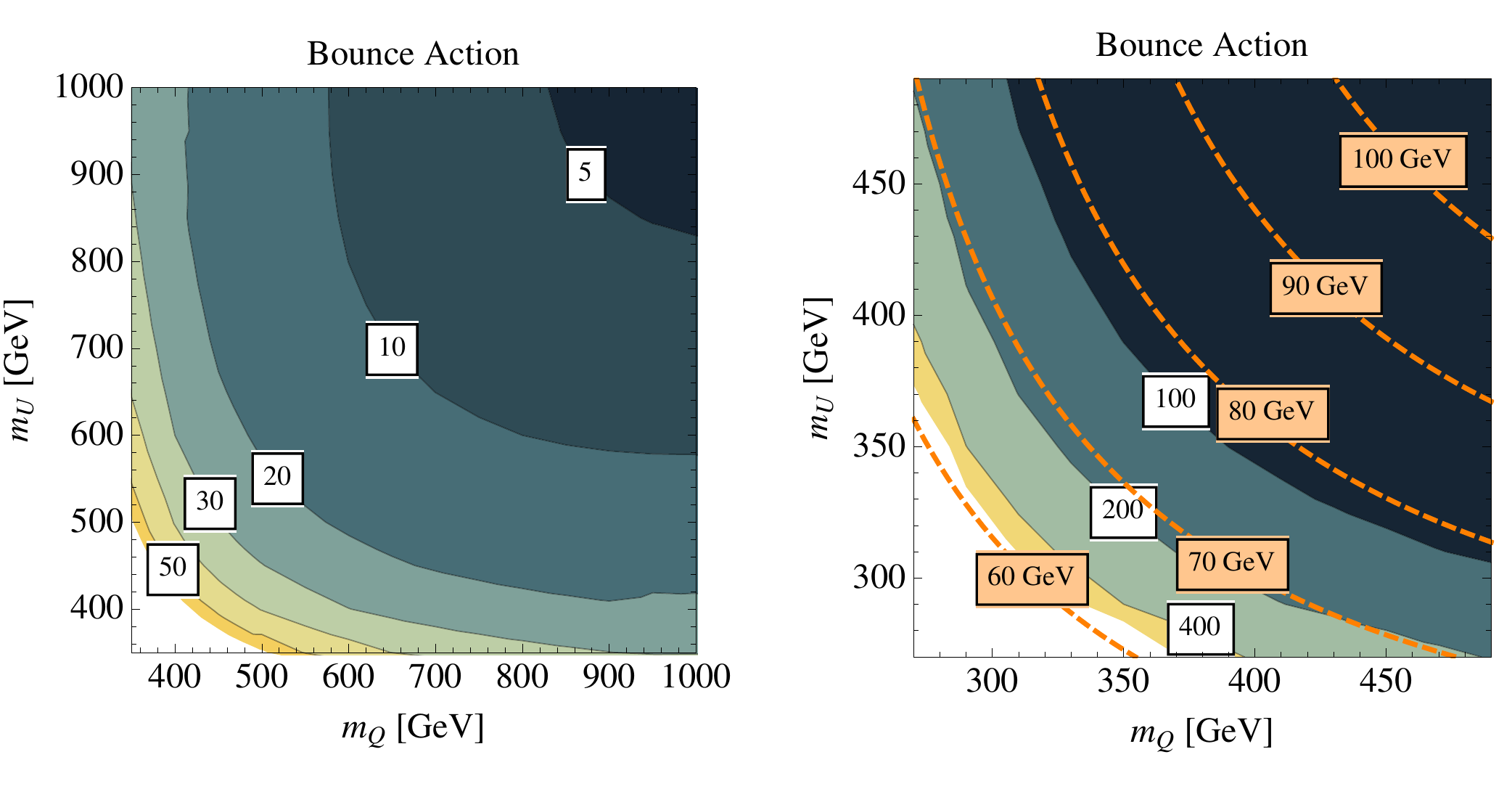}
\end{center}
\caption{Contours of the bounce action $S_0$ as calculated by CosmoTransitions~\cite{Wainwright:2011kj}. The requirement for a sufficiently long-lived vacuum is $S_0 \simgt 400$. The left-hand plot shows that the bulk of the parameter space fails this requirement by a wide margin. The right-hand plot zooms in on the low-mass region, overlaying contours of the mass of the light stop eigenstate ${\tilde t}_1$ (orange, dashed). The bounce action exceeds 400 only when the light stop eigenstate is below 70 GeV, and thus cleanly excluded by LEP constraints.} 
\label{fig:tunnelingrate}
\end{figure}

Because the results of Ref.~\cite{Kusenko:1996jn} are expressed as a scatter plot of points that are viable or not, it is not possible to do a systematic check from their results of whether the parameter space for which the $hGG$ amplitude is inverted (as displayed in Figure~\ref{fig:stopplane}) is ruled out. Thus, we perform a new numerical calculation of the zero-temperature tunneling rate, using a slightly modified version of the CosmoTransitions software~\cite{Wainwright:2011kj}.\footnote{The main change was to replace a call to {\tt scipy.optimize.fmin} with one to {\tt scipy.optimize.fminbound} to prevent a minimum-finding step from skipping over a shallow minimum and falling into a deep one.} The result is depicted in Figure~\ref{fig:tunnelingrate}. In the right-hand panel, one can see that a bounce action $S_0 \simgt 400$, necessary for a sufficiently long-lived metastable vacuum to describe our universe, occurs only for a light stop mass eigenstate below 70 GeV. Such a light stop is excluded by LEP, even in the case of small ${\tilde t}_1-{\tilde \chi}^0_1$ mass splitting~\cite{Abbiendi:2002mp,Achard:2003ge}.

\subsection{Inverting $hGG$ with charged scalar color octets}

Here we will consider a different possibility that does not involve large mixing effects. If we drop the assumption of supersymmetry, we can consider charged scalar octets that have a mass that decreases with increasing Higgs mass,
\beq
V = -\mu^2 H^\dagger H + \lambda_H \left(H^\dagger H\right)^2 + \left(m_O^2 - \lambda_{HO} H^\dagger H\right) O^\dagger O + \lambda_O \left(O^\dagger O\right)^2,
\eeq
with $\lambda_{HO} > 0$. This is a simplified subset of the interactions that arise, for example, for the Manohar-Wise scalar in the $({\bf 8}, {\bf 2})_{1/2}$ representation of the Standard Model gauge group~\cite{Manohar:2006ga}. Other interactions contract the SU(2) indices of $H$ with those of $O$. There is no principled reason to ignore them, but we restrict to a low-dimensional parameter space for ease of plotting the results and because we expect it will capture the qualitative story of the interplay between vacuum stability and Higgs corrections. Quantitatively, it could be worthwhile to explore the full set of operators, but this is beyond the scope of this paper. 

The Manohar-Wise representation contains both a neutral scalar $O^0$ and a charged scalar $O^+$; assuming they have the same mass, as they do with this simplified set of interactions with the Higgs, one finds that they affect the Higgs decay widths as shown by the dashed purple curve in Figure~\ref{fig:higgsoptions}, which comes rather close to the best-fit point of our simplified $\chi^2$ fit. Effects of such an octet scalar on the $hGG$ amplitude were considered recently in Refs.~\cite{Bai:2011aa,Dobrescu:2011aa,Kumar:2012ww} in the regime with relatively small corrections that would lead to a reduced $gg \to H$ cross section. The possibility that $\lambda_{HO} < 0$ could lead to a reasonable fit of the data with enhanced diphoton rate was observed in Ref.~\cite{Batell:2011pz}. Furthermore, as emphasized in Ref.~\cite{Kribs:2012kz}, this regime of parameter space makes a striking prediction of a di-Higgs production rate hundreds or thousands of times larger than the rate in the Standard Model.

\begin{figure}[h]
\begin{center}
\includegraphics[width=0.9\textwidth]{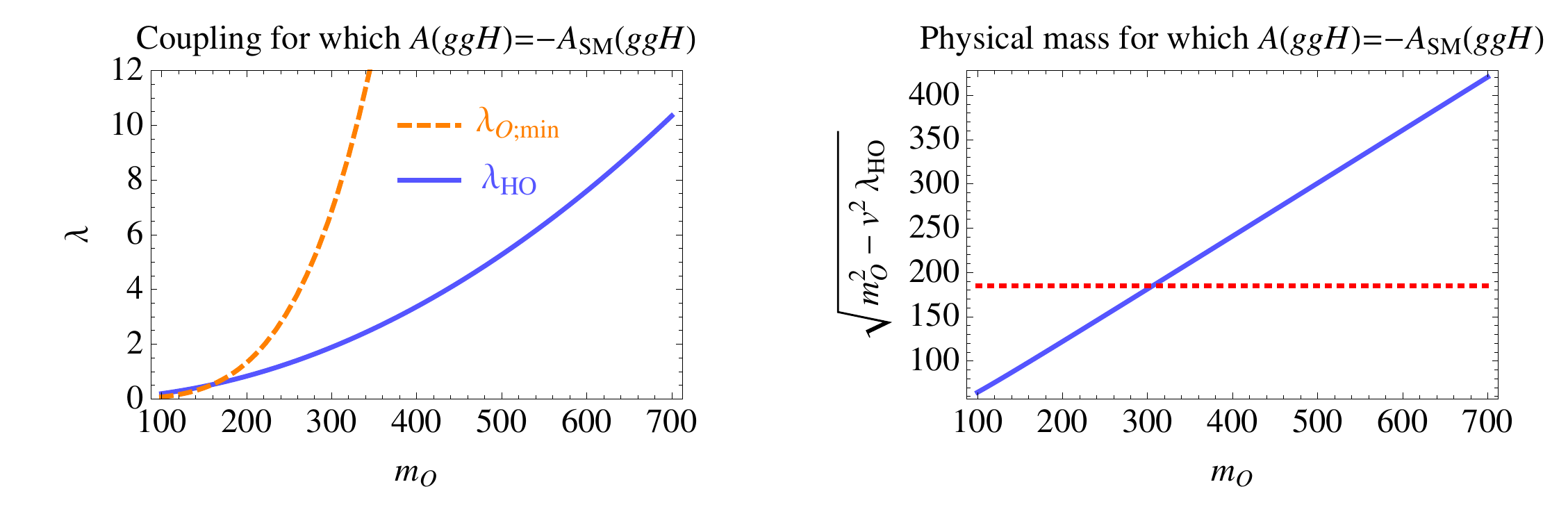}
\end{center}
\caption{Left: value of the Higgs--octet coupling required for a sign-flip of the $hGG$ amplitude (light blue, solid) and of the corresponding minimum octet quartic coupling needed for a potential that is not unbounded below. Right: physical mass of the octet. The dotted red line at 185 GeV marks the lower bound on a sgluon mass from the ATLAS study~\cite{Aad:2011yh}, which may be taken as an approximate guide to the collider constraints on this scenario.} 
\label{fig:octetparams}
\end{figure}

\begin{figure}[h]
\begin{center}
\includegraphics[width=0.42\textwidth]{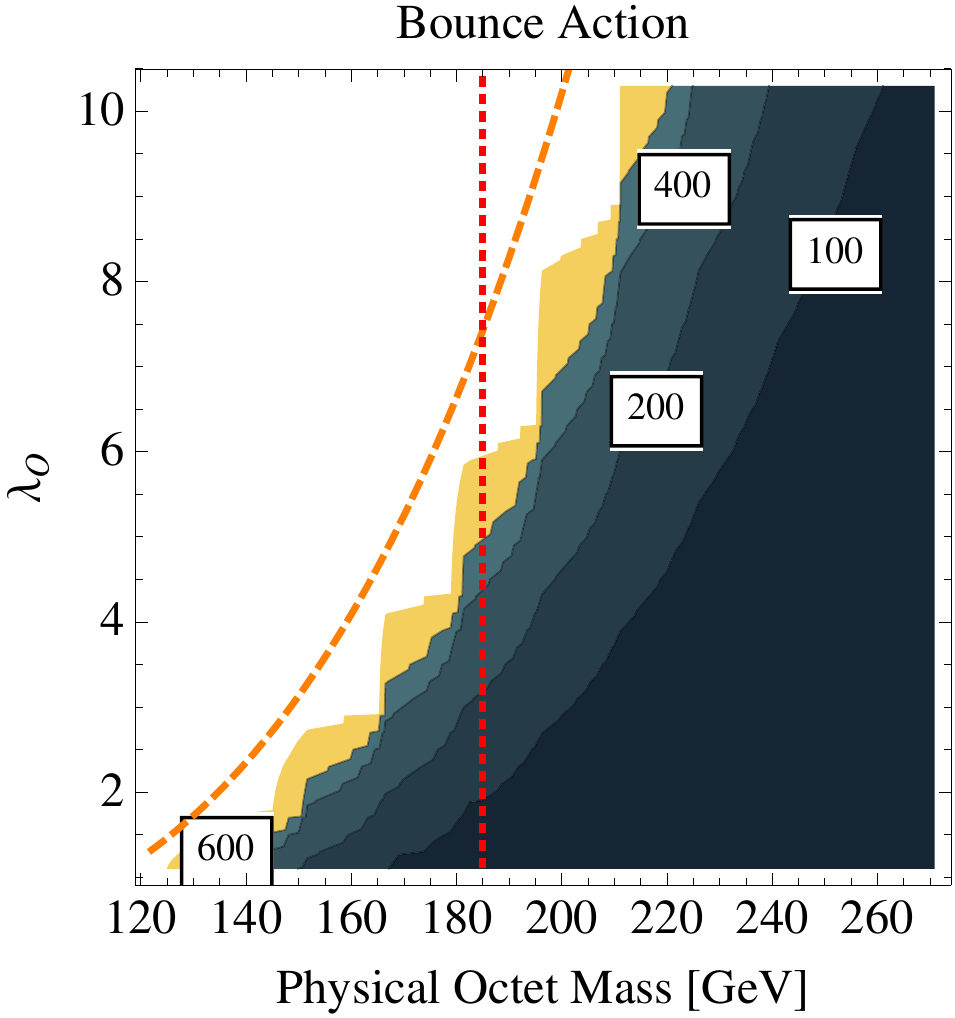}
\end{center}
\caption{The bounce action for tunneling away from the metastable minimum in the scalar octet case, as a function of the physical octet mass $\sqrt{m_O^2 - \lambda_{HO} v^2}$ and the octet quartic $\lambda_O$. The region below the dashed orange curve has a potential that is unbounded from below. Nonetheless, the tunneling calculation shows that a portion of this region is metastable enough to provide a viable vacuum. The vertical red dotted line is an estimate of the collider bound, showing that any surviving parameter space is at masses near 200 GeV and strong coupling $\lambda_O \simgt 4$, or must decay in a manner that evades the ATLAS paired dijet search. Kinks in the curves are from the parameter grid of the numerical scan, not physics.} 
\label{fig:octetbounce}
\end{figure}

In this case, the condition $A_{\rm NP}(hGG) = -2 A_{\rm SM}(hGG)$, at one loop and ignoring $m_O^2/m_H^2$ effects, singles out a particular choice of $\lambda_{HO}$ given the mass $m_O^2$:
\beq
\lambda_{HO} = \frac{16 m_O^2}{25 v^2}.
\eeq
Taking into account the (small) $m_H^2/m_O^2$ corrections, we plot the required choice of $\lambda_{HO}$ as a function of $m_O^2$ in Figure~\ref{fig:octetparams} along with the physical mass of the octet. Notice that, unless the new octet state is very light, the coupling quickly becomes extremely large. In particular, once the physical octet mass reaches about 400 GeV, the coupling is nonperturbatively large. Hence, this scenario is only viable with relatively light states. In fact, the quartic part of the potential becomes unbounded below unless the condition
\beq
\lambda_O \geq \lambda_{O;min} \equiv \frac{\lambda_{HO}^2}{4 \lambda_H}
\eeq
is satisfied. We have also plotted $\lambda_{O;min}$ in Fig.~\ref{fig:octetparams}. It becomes nonperturbatively large already when $m_O \approx 300$ GeV, a point at which the physical mass is only about 180 GeV. Of course, a potential that is unbounded below does not, strictly speaking, exclude the theory; this requires a check of the tunneling rate from our metastable vacuum to the runaway part of the potential, as in the previous section. We show this tunneling rate in Figure~\ref{fig:octetbounce}, which indicates that a value of $\lambda_O$ a factor of 1.5 to 2 below $\lambda_{O;min}$ can yield an unbounded-from-below potential that is metastable enough to be compatible with the age of our universe.

The full Lagrangian of Ref.~\cite{Manohar:2006ga}, including further operators such as $H^{\dagger a}H^b O^{\dagger A}_a O^A_b$ (with $a,b$ SU(2)$_L$ indices and $A$ an SU(3)$_c$ index) and Yukawa couplings of $O$ to SM fermions, is beyond the scope of this paper. Nonetheless, we will make brief remarks on collider bounds. MFV Yukawa couplings of $O$ to the quark fields lead to dominant decays $O^+ \to t{\bar b}$ and $O^0 \to t{\bar t}$ (when this mode is kinematically accessible). However, in most of the mass range that is viable for flipping the $hGG$ amplitude, the decay to tops will be shut off. In that case, the searches for paired dijet resonances performed by ATLAS~\cite{Aad:2011yh} and CMS~\cite{CMS-PAS-EXO-11-016} are likely the most sensitive probes of the scalar octets. (However, depending on the splitting within the SU(2)$_L$ multiplet, searches relying on leptons may also set bounds~\cite{Bai:2011aa}.) The CMS dijet resonance study only constrains states above 320 GeV, due to the relatively hard cuts required by high-luminosity running. The ATLAS study relied on early data with lower trigger thresholds, and bounds sgluons to be heavier than 185 GeV. Because we have multiple octet states, it is possible that the bound is stronger, but this conclusion depends on details of the branching ratios of our octets. Rather than undertake a full study of the collider bounds, we show the 185 GeV bound in Figures~\ref{fig:octetparams} and~\ref{fig:octetbounce} as a rough guideline. This shows that the viable parameter space is in a narrow range of masses above the bound and at strong coupling $\lambda_O \simgt 4$, unless the octet decays in a way that evades the ATLAS search. A more detailed discussion of constraints on Manohar-Wise octet scalars may be found in Ref.~\cite{Burgess:2009wm}. Another recent update on collider bounds is in Ref.~\cite{Altmannshofer:2012ur}.

\subsection{Inverting $hGG$ with new fermions}

Having explored the effects of scalars that change the sign of $hGG$ with large mixing effects or with negative quartics, and shown that there are vacuum stability problems in both cases, we should make some remarks on the case of fermions. Because qualitatively similar observations were made recently in Ref.~\cite{ArkaniHamed:2012kq}, we will be brief. The essential point is that new color triplet fermions with Yukawa couplings to the Higgs contribute terms $\frac{d\lambda}{dt} = -\frac{3}{8\pi^2} y^4$ in the RGE for the Higgs quartic. These corrections drive $\lambda$ negative at relatively low energies, leading to yet another vacuum instability. Of course, there is a way out: if the new colored fields come in complete supermultiplets, the scalars contribute an opposite contribution to the running of $\lambda$ and the quartic can be saved from turning negative. Thus, one perspective on this correction is that it gives a bound on the size of the allowed splitting between fermions and scalars in the new multiplet; this is essentially the naturalness point of view discussed in Ref.~\cite{ArkaniHamed:2012kq}.

The first observation relates to fermionic top partners. In particular, suppose we have new fields $T, {\overline T}$ in the $({\bf 3},{\bf 1})_{\pm 2/3}$ representations of the Standard Model gauge group. We can add both a vectorlike mass for these fields and a mixing term with the SM left-handed quarks,
\beq
M T {\bar T} + y_T H Q {\overline T} + y_{\overline T} H^\dagger Q T.
\eeq
Such top partners contribute a correction to the $hGG$ amplitude:
\beq
\frac{A(hGG)}{A_{\rm SM}(hGG)} = 1 - \frac{v y_T y_{\overline T}}{M y_t - v y_T y_{\overline T}}.
\eeq
If we wish this to equal $-1$, we must take $y_T y_{\overline T} = \frac{2}{3}y_t \frac{M}{v}$. If the new colored states are to be heavier than the top quark, this requires large Yukawas. Furthermore, these states are highly mixed with the top, and require that we significantly alter $y_t$ from its Standard Model value. This is an awkward solution that will be difficult to reconcile with experimental bounds.

A safer approach is to add a pair of vectorlike fermions, as in Section~\ref{sec:fermions}, which are not mixed with the SM top. To be concrete, we will take these states to have the same quantum numbers as the SM $Q$ and $u^c$ fields, but with a parity that prevents mixing terms with the SM. Furthermore, we will simplify the story by taking $m_\chi = m_\psi = M$ and $y_1 = y_2 = y$. Obtaining the amplitude $A(hGG) = - A_{\rm SM}(hGG)$ then requires that $y^2 v^2 = \frac{1}{2} M^2$, with mass eigenvalues about $M_{\rm light} \approx 0.29 M$ and $M_{\rm heavy} \approx 1.7 M$. The finite correction to $\lambda$ (which must have a value of about $0.13$ for the correct Higgs VEV) from the Coleman-Weinberg formula can then be expressed as $-\frac{M^4}{4\pi^2 v^4} \approx -3.4 \frac{M_{\rm light}^4}{v^4}$, so that as we raise the mass scale of the new colored fermions relative to the top mass, the tuning in the Higgs sector increases quartically.

\begin{figure}[h]
\begin{center}
\includegraphics[width=0.45\textwidth]{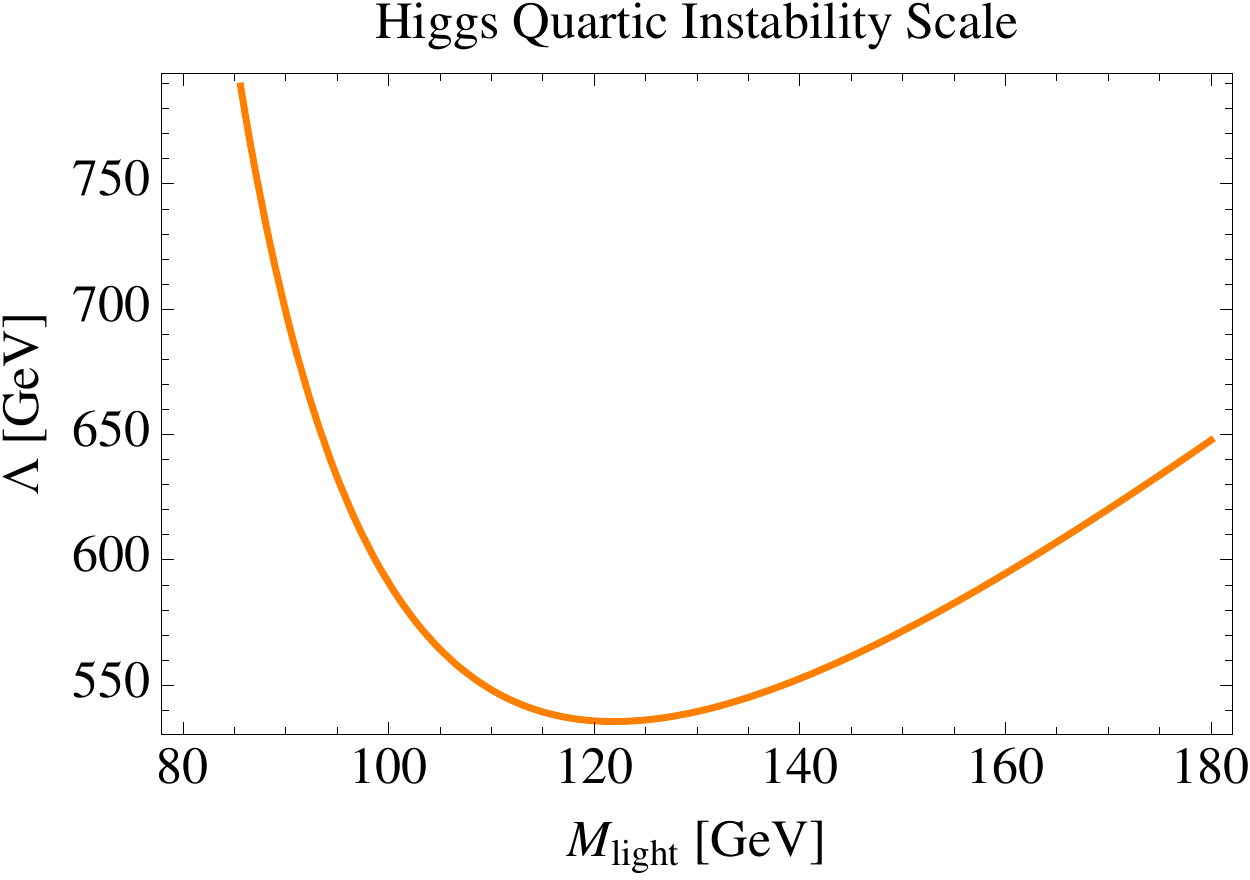}
\end{center}
\caption{An approximation to the scale $\Lambda$ at which an instability in the Higgs potential sets in, as a function of the light fermion mass eigenstate $M_{\rm light}$. See the text for an explanation.} 
\label{fig:fermioninstability}
\end{figure}

Finally, we give an approximate solution of the RGEs to see at what scale $\Lambda$ the Higgs quartic drives the potential unstable, when $\lambda(\Lambda) = \frac{2\pi^2}{3\log(H/\Lambda)}$, as in Ref.~\cite{ArkaniHamed:2012kq}. For simplicity we have dropped terms in the RGEs proportional to $g_1$, which do not significantly change the results. We begin at the $\overline{MS}$ top mass in the Standard Model, run up to the scale $M$ using Standard Model beta functions, and then run to higher energies with the new physics beta functions, turning on $y_1 = y_2$ at $M$. The result is shown in Figure~\ref{fig:fermioninstability}. The rising curve at $M_{\rm light} \simgt 120$ GeV approximately tracks the value of $M \approx 3 M_{\rm light}$, indicating that $\lambda$ runs negative essentially immediately when we turn on the RG effects of the new states. A better calculation would correctly take into account the running between the thresholds $M_{\rm light}$ and $M_{\rm heavy}$, but this plot makes our qualitative point: if new fermionic states are to change the sign of the $hGG$ amplitude, not only do they imply an uncomfortably large amount of fine-tuning and strong coupling, but their superpartners {\em must} be nearby. Otherwise, they are ruled out by a catastrophic vacuum instability, much like the scalar cases we have studied.

\section{Discussion}
\label{sec:discuss}

We have seen that, in any region with large enough radiative corrections from loops of new colored and charged particles to flip the sign of the $hGG$ amplitude, there are significant modifications to the Higgs potential and potentially dangerous radiative effects. In particular, the most appealing such scenario, with loops of stop squarks, is ruled out by rapid vacuum decay to color- and charge-breaking minima. In the case of a color octet scalar with a negative quartic coupling to the Higgs, the combination of vacuum decay bounds and collider constraints rules out much of the parameter space. However, a light octet scalar around 200 GeV with a large self-coupling may still be allowed. This loophole could likely be closed by a more thorough analysis, or by further collider searches. Fermionic states are only allowed if they are part of a supermultiplet with the scalar states nearby.

In the scalar cases, one could ask whether adding new terms to the potential, beyond those we have considered, could lift the dangerous minima and render the $A(hGG) = -A_{\rm SM}(hGG)$ scenario viable after all. However, a local change in the potential far from the good EWSB vacuum is unlikely to have much effect, since in the stop case the tunneling is to a very deep minimum, and in the octet scalar case to a runaway direction. In both scenarios, the fundamental problem is that a relatively low barrier separates the vacuum that could represent our universe from a steep downhill plunge. Any physics that could make this viable has to change the potential {\em near} our vacuum, making the shallow hill in the potential into a sizable barrier. This likely requires new strong coupling, and although such models would have to be analyzed on a case-by-case basis, it seems unlikely that a model that could achieve this would not also alter Higgs production or decay in other ways, rendering the original motivation moot.

A safer scenario to fit possible deviations in the data is to rely on loop corrections of charged color-singlet particles to enhance the $h\gamma\gamma$ rate. This has received attention recently in Refs.~\cite{Carena:2011aa,Carena:2012gp,Carena:2012xa,Joglekar:2012hb,ArkaniHamed:2012kq,An:2012vp,Almeida:2012bq}. In the scenarios involving new scalars, it may be worthwhile to do a careful scan for charge-violating minima and tunneling rates that could constrain the parameter space in a similar way to that we discussed here. The various difficulties with tuning and vacuum instabilities arise simply because achieving large effects with loops requires venturing into extreme regions of parameter space. (A distinctive scenario in which the correct sign of the amplitude arises is from loops of new charged gauge bosons~\cite{Alves:2011kc,Carena:2012xa,Alves:2012yp}.) If the LHC observations continue to indicate substantial deviations in Higgs properties, it may mean that the effect arises at tree-level, which is easily achieved by non-decoupling effects of further Higgs states~\cite{Azatov:2012wq,Hall:2011aa,Blum:2012kn,Hagiwara:2012mg,Bellazzini:2012mh,Craig:2012vn,Alves:2012ez}. Searching for such states should continue to be a central part of the LHC's ongoing investigation of the nature of electroweak symmetry breaking.

\section*{Acknowledgments}
I thank Patrick Meade for useful discussions that helped to initiate this project, and Max Wainwright for helpful correspondence regarding the CosmoTransitions software. I also thank Haipeng An, JiJi Fan, Christophe Grojean, Veronica Sanz, and Mike Trott for interesting discussions or correspondence, and Mike Trott for comments on the draft. A portion of this work was carried out while visiting the Perimeter Institute for Theoretical Physics. Research at Perimeter Institute is supported by the Government of Canada through Industry Canada and by the Province of Ontario through the Ministry of Economic Development \& Innovation. This work was also supported in part by the Fundamental Laws Initiative of the Harvard Center for the Fundamental Laws of Nature.

\appendix

\section{Details of the Coleman-Weinberg calculations}
\label{app:CW}

Here we present the formulas for the Coleman-Weinberg corrections to the quartic. First, the case of fermions discussed in Section~\ref{sec:fermions} gives:
\beq
\delta \lambda_F =  && -\frac{N_{c;F}}{16\pi^2}\left\{\frac{2\left[y_1 y_2 \left(m_\chi^2 + m_\psi^2\right) + \left(y_1^2 + y_2^2\right) m_\chi m_\psi\right]^2}{\left(m_\chi^2 - m_\psi^2\right)^2} \right.  \nonumber \\
& & + \log\frac{m_\chi^2}{\mu^2}  \left.m_\chi^2 \frac{\left(y_1^4 + y_2^4\right) m_\chi^4 -3 \left(y_1^2 + y_2^2\right)^2 m_\chi^2 m_\psi^2 - 8 y_1 y_2\left(y_1^2+y_2^2\right) m_\chi m_\psi^3- 6 y_1^2 y_2^2 m_\psi^4}{\left(m_\chi^2 - m_\psi^2\right)^3} \right.\nonumber \\
&& - \log\frac{m_\psi^2}{\mu^2}  \left. m_\psi^2 \frac{\left(y_1^4 + y_2^4\right) m_\psi^4 -3 \left(y_1^2 + y_2^2\right)^2 m_\psi^2 m_\chi^2 - 8 y_1 y_2 \left(y_1^2 + y_2^2\right) m_\psi m_\chi^3 - 6 y_1^2 y_2^2 m_\chi^4}{\left(m_\chi^2 - m_\psi^2\right)^3} \right\} \nonumber \\
 \xrightarrow{m_\chi = m_\psi = m} && -\frac{N_{c;F}}{16\pi^2} \left\{\left(y_1^4 + y_2^4\right) \log \frac{m^2}{\mu^2} + \frac{1}{6}\left(5y_1^2 -2 y_1 y_2 + 5y_2^2\right)\left(y_1+y_2\right)^2 \right\}.
\eeq
The case of scalars discussed in Section~\ref{sec:scalars} gives:
\beq
\delta \lambda_S & = & \frac{N_{c;S}}{32\pi^2} \left\{\frac{2 A^2\left(A^2 + \left(m_1^2-m_2^2\right)\left(\lambda_1-\lambda_2\right)\right)}{\left(m_1^2 - m_2^2\right)^2} \right. \nonumber \\
&  & + \log\frac{m_2^2}{\mu^2} \left(\frac{A^4 \left(m_1^2 + m_2^2\right)}{\left(m_1^2 - m_2^2\right)^3} - \frac{2A^2 \left(m_1^2 \lambda_2 - m_2^2 \lambda_1\right)}{\left(m_1^2 - m_2^2\right)^2} + \lambda_2^2\right) \nonumber \\
&  & \left. + \log\frac{m_1^2}{\mu^2} \left(\frac{A^4 \left(m_1^2 + m_2^2\right)}{\left(m_1^2 - m_2^2\right)^3} - \frac{2A^2 \left(m_1^2 \lambda_2 - m_2^2 \lambda_1\right)}{\left(m_1^2 - m_2^2\right)^2} - \lambda_1^2\right)\right\} \\
& \xrightarrow{m_1 = m_2 = m} & \frac{N_{c;S}}{32\pi^2} \left((\lambda_1^2 + \lambda_2^2)\log\frac{m^2}{\mu^2} + (\lambda_1 + \lambda_2) \frac{A^2}{m^2} - \frac{1}{6} \frac{A^4}{m^4}\right).
\eeq


\begin{thebibliography}{999}

\bibitem{ATLAS:2012gk} 
  G.~Aad {\it et al.}  [ATLAS Collaboration],
  ``Observation of a new particle in the search for the Standard Model Higgs boson with the ATLAS detector at the LHC,''
  \href{http://arxiv.org/abs/arXiv:1207.7214}{arXiv:1207.7214 [hep-ex]}.

\bibitem{CMS:2012gu} 
  S.~Chatrchyan {\it et al.}  [CMS Collaboration],
  ``Observation of a new boson at a mass of 125 GeV with the CMS experiment at the LHC,''
  \href{http://arxiv.org/abs/arXiv:1207.7235}{arXiv:1207.7235 [hep-ex]}.
  
\bibitem{CMS-PAS-HIG-12-015}
 The CMS Collaboration,
 ``Evidence for a new state decaying into two photons in the search for the standard model Higgs boson in pp collisions,''
  \href{http://cdsweb.cern.ch/record/1460419}{CMS-PAS-HIG-12-015}.

\bibitem{ATLAS-CONF-2012-091}
 The ATLAS Collaboration,
 ``Observation of an excess of events in the search for the Standard Model Higgs boson in the gamma-gamma channel with the ATLAS detector,''
  \href{http://cdsweb.cern.ch/record/1460410}{ATLAS-CONF-2012-091}.  

\bibitem{Englert:2011aa} 
  C.~Englert, T.~Plehn, M.~Rauch, D.~Zerwas and P.~M.~Zerwas,
  ``LHC: Standard Higgs and Hidden Higgs,''
  Phys.\ Lett.\ B {\bf 707}, 512 (2012)
  \href{http://arxiv.org/abs/arXiv:1112.3007}{arXiv:1112.3007 [hep-ph]}.

\bibitem{Carmi:2012yp} 
  D.~Carmi, A.~Falkowski, E.~Kuflik and T.~Volansky,
  ``Interpreting LHC Higgs Results from Natural New Physics Perspective,''
  \href{http://arxiv.org/abs/1202.3144}{arXiv:1202.3144 [hep-ph]}.
  
\bibitem{Azatov:2012bz} 
  A.~Azatov, R.~Contino and J.~Galloway,
  ``Model-Independent Bounds on a Light Higgs,''
  JHEP {\bf 1204}, 127 (2012)
  \href{http://arxiv.org/abs/1202.3415}{arXiv:1202.3415 [hep-ph]}.
  
\bibitem{Espinosa:2012ir} 
  J.~R.~Espinosa, C.~Grojean, M.~Muhlleitner and M.~Trott,
  ``Fingerprinting Higgs Suspects at the LHC,''
  JHEP {\bf 1205}, 097 (2012)
  \href{http://arxiv.org/abs/1202.3697}{arXiv:1202.3697 [hep-ph]}.

\bibitem{Giardino:2012ww} 
  P.~P.~Giardino, K.~Kannike, M.~Raidal and A.~Strumia,
  ``Reconstructing Higgs boson properties from the LHC and Tevatron data,''
  JHEP {\bf 1206}, 117 (2012)
  \href{http://arxiv.org/abs/1203.4254}{arXiv:1203.4254 [hep-ph]}.
  
\bibitem{Li:2012ku} 
  T.~Li, X.~Wan, Y.-K.~Wang and S.-H.~Zhu,
  ``Constraints on the Universal Varying Yukawa Couplings: from SM-like to Fermiophobic,''
  \href{http://arxiv.org/abs/1203.5083}{arXiv:1203.5083 [hep-ph]}.
  
\bibitem{Rauch:2012wa} 
  M.~Rauch,
  ``Determination of Higgs-boson couplings (SFitter),''
  \href{http://arxiv.org/abs/1203.6826}{arXiv:1203.6826 [hep-ph]}.
  
\bibitem{Ellis:2012rx} 
  J.~Ellis and T.~You,
  ``Global Analysis of Experimental Constraints on a Possible Higgs-Like Particle with Mass ~ 125 GeV,''
  JHEP {\bf 1206}, 140 (2012)
  \href{http://arxiv.org/abs/1204.0464}{arXiv:1204.0464 [hep-ph]}.

\bibitem{Klute:2012pu} 
  M.~Klute, R.~Lafaye, T.~Plehn, M.~Rauch and D.~Zerwas,
  ``Measuring Higgs Couplings from LHC Data,''
  Phys.\ Rev.\ Lett.\  {\bf 109}, 101801 (2012)
  \href{http://arxiv.org/abs/arXiv:1205.2699}{arXiv:1205.2699 [hep-ph]}.
  
\bibitem{Azatov:2012wq} 
  A.~Azatov, S.~Chang, N.~Craig and J.~Galloway,
  ``Early Higgs Hints for Non-Minimal Supersymmetry,''
  \href{http://arxiv.org/abs/1206.1058}{arXiv:1206.1058 [hep-ph]}.
    
\bibitem{Carmi:2012zd} 
  D.~Carmi, A.~Falkowski, E.~Kuflik and T.~Volansky,
  ``Interpreting the Higgs,''
  \href{http://arxiv.org/abs/1206.4201}{arXiv:1206.4201 [hep-ph]}.

\bibitem{Corbett:2012dm} 
  T.~Corbett, O.~J.~P.~Eboli, J.~Gonzalez-Fraile and M.~C.~Gonzalez-Garcia,
  ``Constraining anomalous Higgs interactions,''
  \href{http://arxiv.org/abs/arXiv:1207.1344}{arXiv:1207.1344 [hep-ph]}.

\bibitem{Giardino:2012dp} 
  P.~P.~Giardino, K.~Kannike, M.~Raidal and A.~Strumia,
  ``Is the resonance at 125 GeV the Higgs boson?,''
  \href{http://arxiv.org/abs/1207.1347}{arXiv:1207.1347 [hep-ph]}.
  
\bibitem{Buckley:2012em} 
  M.~R.~Buckley and D.~Hooper,
  ``Are There Hints of Light Stops in Recent Higgs Search Results?,''
  \href{http://arxiv.org/abs/1207.1445}{arXiv:1207.1445 [hep-ph]}.
  
\bibitem{Ellis:2012hz} 
  J.~Ellis and T.~You,
  ``Global Analysis of the Higgs Candidate with Mass $\sim$ 125 GeV,''
  \href{http://arxiv.org/abs/1207.1693}{arXiv:1207.1693 [hep-ph]}.
  
\bibitem{Montull:2012ik} 
  M.~Montull and F.~Riva,
  ``Higgs discovery: the beginning or the end of natural EWSB?,''
  \href{http://arxiv.org/abs/arXiv:1207.1716}{arXiv:1207.1716 [hep-ph]}.
  
\bibitem{Espinosa:2012im} 
  J.~R.~Espinosa, C.~Grojean, M.~Muhlleitner and M.~Trott,
  ``First Glimpses at Higgs' face,''
  \href{http://arxiv.org/abs/1207.1717}{arXiv:1207.1717 [hep-ph]}.
  
\bibitem{Carmi:2012in} 
  D.~Carmi, A.~Falkowski, E.~Kuflik, T.~Volansky and J.~Zupan,
  ``Higgs After the Discovery: A Status Report,''
  \href{http://arxiv.org/abs/1207.1718}{arXiv:1207.1718 [hep-ph]}.
  
\bibitem{Plehn:2012iz} 
  T.~Plehn and M.~Rauch,
  ``Higgs Couplings after the Discovery,''
  \href{http://arxiv.org/abs/arXiv:1207.6108}{arXiv:1207.6108 [hep-ph]}.
  
\bibitem{Espinosa:2012in} 
  J.~R.~Espinosa, C.~Grojean, V.~Sanz and M.~Trott,
  ``NSUSY fits,''
  \href{http://arxiv.org/abs/arXiv:1207.7355}{arXiv:1207.7355 [hep-ph]}.
  
\bibitem{ATLAS-CONF-2012-092}
 The ATLAS Collaboration,
 ``Observation of an excess of events in the search for the Standard Model Higgs boson in the $H \to ZZ^{(*)} \to 4 \ell$ channel with the ATLAS detector,''
  \href{http://cdsweb.cern.ch/record/1460411}{ATLAS-CONF-2012-092}. 
  
\bibitem{CMS-PAS-HIG-12-016}
 The CMS Collaboration,
 ``Evidence for a new state in the search for the standard model Higgs boson in the H to ZZ to 4 leptons channel in pp collisions at sqrt(s) = 7 and 8 TeV,''
  \href{http://cdsweb.cern.ch/record/1460664}{CMS-PAS-HIG-12-016}.   

\bibitem{CMS-PAS-HIG-12-017}
 The CMS Collaboration,
 ``Search for the standard model Higgs boson decaying to a W pair in the fully leptonic final state in pp collisions at sqrt(s) = 8 TeV,''
  \href{http://cdsweb.cern.ch/record/1460424}{CMS-PAS-HIG-12-017}.
  
\bibitem{ATLAS-CONF-2012-098}
 The ATLAS Collaboration,
 ``Observation of an Excess of Events in the Search for the Standard Model Higgs Boson in the $H \to WW^{(*)} \to \ell \nu \ell \nu$ Channel with the ATLAS Detector,''
  \href{http://cdsweb.cern.ch/record/1462530}{ATLAS-CONF-2012-098}.  

\bibitem{Dimopoulos:1995mi} 
  S.~Dimopoulos and G.~F.~Giudice,
  ``Naturalness constraints in supersymmetric theories with nonuniversal soft terms,''
  Phys.\ Lett.\ B {\bf 357}, 573 (1995)
  \href{http://arxiv.org/abs/hep-ph/9507282}{hep-ph/9507282}.
 
\bibitem{Cohen:1996vb} 
  A.~G.~Cohen, D.~B.~Kaplan and A.~E.~Nelson,
  ``The More minimal supersymmetric standard model,''
  Phys.\ Lett.\ B {\bf 388}, 588 (1996)
  \href{http://arxiv.org/abs/hep-ph/9607394}{hep-ph/9607394}.
  
\bibitem{Ellis:1975ap} 
  J.~R.~Ellis, M.~K.~Gaillard and D.~V.~Nanopoulos,
  ``A Phenomenological Profile of the Higgs Boson,''
  \href{http://dx.doi.org/10.1016/0550-3213(76)90382-5}{Nucl.\ Phys.\ B {\bf 106}, 292 (1976)}.
  
\bibitem{Shifman:1979eb} 
  M.~A.~Shifman, A.~I.~Vainshtein, M.~B.~Voloshin and V.~I.~Zakharov,
  ``Low-Energy Theorems for Higgs Boson Couplings to Photons,''
  \href{http://inspirehep.net/record/141287?ln=en}{Sov.\ J.\ Nucl.\ Phys.\  {\bf 30}, 711 (1979)}
  [Yad.\ Fiz.\  {\bf 30}, 1368 (1979)].
  
\bibitem{Cohen:2012wg} 
  A.~G.~Cohen and M.~Schmaltz,
  ``New Charged Particles from Higgs Couplings,''
  \href{http://arxiv.org/abs/1207.3495}{arXiv:1207.3495 [hep-ph]}.
  
\bibitem{Dorsner:2012new}
  I.~Dorsner, S.~Fajfer, A.~Greljo, and J.~F.~Kamenik,
  ``Higgs Uncovering Light Scalar Remnants of High Scale Matter Unification,''
  \href{http://arxiv.org/abs/1208.1266}{arXiv:1208.1266 [hep-ph]}.
  
\bibitem{Martin:1997ns} 
  S.~P.~Martin,
  ``A Supersymmetry primer,''
  In Kane, G.L. (ed.): Perspectives on supersymmetry II 1-153
  \href{http://arxiv.org/abs/hep-ph/9709356}{hep-ph/9709356}.
  
\bibitem{Carena:1995wu} 
  M.~S.~Carena, M.~Quiros and C.~E.~M.~Wagner,
  ``Effective potential methods and the Higgs mass spectrum in the MSSM,''
  Nucl.\ Phys.\ B {\bf 461}, 407 (1996)
  \href{http://arxiv.org/abs/hep-ph/9508343}{hep-ph/9508343}.
  
\bibitem{ArkaniHamed:2012kq} 
  N.~Arkani-Hamed, K.~Blum, R.~T.~D'Agnolo and J.~Fan,
  ``2:1 for Naturalness at the LHC?,''
  \href{http://arxiv.org/abs/1207.4482}{arXiv:1207.4482 [hep-ph]}.
  
\bibitem{Kusenko:1996jn} 
  A.~Kusenko, P.~Langacker and G.~Segre,
  ``Phase transitions and vacuum tunneling into charge and color breaking minima in the MSSM,''
  Phys.\ Rev.\ D {\bf 54}, 5824 (1996)
  \href{http://arxiv.org/abs/hep-ph/9602414}{hep-ph/9602414}.
  
\bibitem{Blum:2012ii} 
  K.~Blum, R.~T.~D'Agnolo and J.~Fan,
  ``Natural SUSY Predicts: Higgs Couplings,''
  \href{http://arxiv.org/abs/arXiv:1206.5303}{arXiv:1206.5303 [hep-ph]}.
  
\bibitem{ATLAS-CONF-2012-070}
 The ATLAS Collaboration,
 ``Search for light top squark pair production in final states with leptons and b-jets with the ATLAS detector in $\sqrt{s}$ = 7 TeV proton--proton collisions,''
  \href{http://cdsweb.cern.ch/record/1460267}{ATLAS-CONF-2012-070}.    

\bibitem{ATLAS-CONF-2012-071}
 The ATLAS Collaboration,
 ``Search for light top squark pair production in final states with leptons and b-jets with the ATLAS detector in $\sqrt{s}$ = 7 TeV proton--proton collisions,''
  \href{http://cdsweb.cern.ch/record/1460268}{ATLAS-CONF-2012-071}.    

\bibitem{ATLAS-CONF-2012-073}
 The ATLAS Collaboration,
 ``Search for direct top squark pair production in final states with one isolated lepton, jets, and missing transverse momentum in $\sqrt{s}$ = 7 TeV pp collisions using 4.7 fb$^{-1}$ of ATLAS data ,''
  \href{http://cdsweb.cern.ch/record/1460270}{ATLAS-CONF-2012-073}.    

\bibitem{ATLAS-CONF-2012-074}
 The ATLAS Collaboration,
 ``Search for a supersymmetric partner of the top quark in final states with jets and missing transverse momentum at $\sqrt{s}$=7 TeV with the ATLAS detector ,''
  \href{http://cdsweb.cern.ch/record/1460271}{ATLAS-CONF-2012-074}.    

\bibitem{CMS-PAS-SUS-12-009}
 The CMS Collaboration,
 ``A search for the decays of a new heavy particle in multijet events with the razor variables at CMS in pp collisions at $\sqrt{s}$=7 TeV,''
   \href{http://cdsweb.cern.ch/record/1459812}{CMS-PAS-SUS-12-009}.
   
\bibitem{CMS-PAS-SUS-11-022}
 The CMS Collaboration,
  ``Search for supersymmetery in final states with missing transverse momentum and 0, 1, 2, or $\geq$ 3 $b$ jets with CMS,''
   \href{http://cdsweb.cern.ch/record/1461947}{CMS-PAS-SUS-11-022}.
   
\bibitem{Meade:2006dw} 
  P.~Meade and M.~Reece,
  ``Top partners at the LHC: Spin and mass measurement,''
  Phys.\ Rev.\ D {\bf 74}, 015010 (2006)
  \href{http://arxiv.org/abs/hep-ph/0601124}{hep-ph/0601124}.

\bibitem{Han:2008gy} 
  T.~Han, R.~Mahbubani, D.~G.~E.~Walker and L.~-T.~Wang,
  ``Top Quark Pair plus Large Missing Energy at the LHC,''
  JHEP {\bf 0905}, 117 (2009)
  \href{http://arxiv.org/abs/arXiv:0803.3820}{arXiv:0803.3820 [hep-ph]}.
  
\bibitem{Plehn:2010st} 
  T.~Plehn, M.~Spannowsky, M.~Takeuchi and D.~Zerwas,
  ``Stop Reconstruction with Tagged Tops,''
  JHEP {\bf 1010}, 078 (2010)
  \href{http://arxiv.org/abs/arXiv:1006.2833}{arXiv:1006.2833 [hep-ph]}.
  
\bibitem{Asano:2010ut} 
  M.~Asano, H.~D.~Kim, R.~Kitano and Y.~Shimizu,
  ``Natural Supersymmetry at the LHC,''
  JHEP {\bf 1012}, 019 (2010)
  \href{http://arxiv.org/abs/arXiv:1010.0692}{arXiv:1010.0692 [hep-ph]}.

\bibitem{Plehn:2011tf} 
  T.~Plehn, M.~Spannowsky and M.~Takeuchi,
  ``Boosted Semileptonic Tops in Stop Decays,''
  JHEP {\bf 1105}, 135 (2011)
  \href{http://arxiv.org/abs/arXiv:1102.0557}{arXiv:1102.0557 [hep-ph]}.

\bibitem{Bai:2012gs} 
  Y.~Bai, H.~-C.~Cheng, J.~Gallicchio and J.~Gu,
  ``Stop the Top Background of the Stop Search,''
  JHEP {\bf 1207}, 110 (2012)
  \href{http://arxiv.org/abs/arXiv:1203.4813}{arXiv:1203.4813 [hep-ph]}.
  
\bibitem{Plehn:2012pr} 
  T.~Plehn, M.~Spannowsky and M.~Takeuchi,
  ``Stop searches in 2012,''
  \href{http://arxiv.org/abs/arXiv:1205.2696}{arXiv:1205.2696 [hep-ph]}.
  
\bibitem{Alves:2012ft} 
  D.~S.~M.~Alves, M.~R.~Buckley, P.~J.~Fox, J.~D.~Lykken and C.~-T.~Yu,
  ``Stops and MET: the shape of things to come,''
  \href{http://arxiv.org/abs/arXiv:1205.5805}{arXiv:1205.5805 [hep-ph]}.

\bibitem{Kaplan:2012gd} 
  D.~E.~Kaplan, K.~Rehermann and D.~Stolarski,
  ``Searching for Direct Stop Production in Hadronic Top Data at the LHC,''
  JHEP {\bf 1207}, 119 (2012)
  \href{http://arxiv.org/abs/arXiv:1205.5816}{arXiv:1205.5816 [hep-ph]}.

\bibitem{Chen:2012uw} 
  C.~-Y.~Chen, A.~Freitas, T.~Han and K.~S.~M.~Lee,
  ``New Physics from the Top at the LHC,''
  \href{http://arxiv.org/abs/arXiv:1207.4794}{arXiv:1207.4794 [hep-ph]}.
   
\bibitem{Han:2012fw} 
  Z.~Han, A.~Katz, D.~Krohn and M.~Reece,
  ``(Light) Stop Signs,''
  to appear in JHEP,
  \href{http://arxiv.org/abs/arXiv:1205.5808}{arXiv:1205.5808 [hep-ph]}.
   
\bibitem{Brust:2012uf} 
  C.~Brust, A.~Katz and R.~Sundrum,
  ``SUSY Stops at a Bump,''
  \href{http://arxiv.org/abs/arXiv:1206.2353}{arXiv:1206.2353 [hep-ph]}.
  
\bibitem{Wainwright:2011kj} 
  C.~L.~Wainwright,
  ``CosmoTransitions: Computing Cosmological Phase Transition Temperatures and Bubble Profiles with Multiple Fields,''
  Comput.\ Phys.\ Commun.\  {\bf 183}, 2006 (2012)
  \href{http://arxiv.org/abs/1109.4189}{arXiv:1109.4189 [hep-ph]}; Software version 1.0.1 from \href{http://chasm.ucsc.edu/cosmotransitions/}{http://chasm.ucsc.edu/cosmotransitions/}.
  
\bibitem{Abbiendi:2002mp} 
  G.~Abbiendi {\it et al.}  [OPAL Collaboration],
  ``Search for scalar top and scalar bottom quarks at LEP,''
  Phys.\ Lett.\ B {\bf 545}, 272 (2002)
  [Erratum-ibid.\ B {\bf 548}, 258 (2002)]
  \href{http://arxiv.org/abs/hep-ex/0209026}{hep-ex/0209026}.
  
\bibitem{Achard:2003ge} 
  P.~Achard {\it et al.}  [L3 Collaboration],
  ``Search for scalar leptons and scalar quarks at LEP,''
  Phys.\ Lett.\ B {\bf 580}, 37 (2004)
  \href{http://arxiv.org/abs/hep-ex/0310007}{hep-ex/0310007}.

\bibitem{Manohar:2006ga} 
  A.~V.~Manohar and M.~B.~Wise,
  ``Flavor changing neutral currents, an extended scalar sector, and the Higgs production rate at the CERN LHC,''
  Phys.\ Rev.\ D {\bf 74}, 035009 (2006)
  \href{http://arxiv.org/abs/hep-ph/0606172}{hep-ph/0606172}.
  
\bibitem{Bai:2011aa} 
  Y.~Bai, J.~Fan and J.~L.~Hewett,
  ``Hiding a Heavy Higgs Boson at the 7 TeV LHC,''
  \href{http://arxiv.org/abs/arXiv:1112.1964}{arXiv:1112.1964 [hep-ph]}.
  
\bibitem{Dobrescu:2011aa} 
  B.~A.~Dobrescu, G.~D.~Kribs and A.~Martin,
  ``Higgs Underproduction at the LHC,''
  Phys.\ Rev.\ D {\bf 85}, 074031 (2012)
  \href{http://arxiv.org/abs/arXiv:1112.2208}{arXiv:1112.2208 [hep-ph]}.
  
\bibitem{Kumar:2012ww} 
  K.~Kumar, R.~Vega-Morales and F.~Yu,
  ``Effects from New Colored States and the Higgs Portal on Gluon Fusion and Higgs Decays,''
  \href{http://arxiv.org/abs/arXiv:1205.4244}{arXiv:1205.4244 [hep-ph]}.
  
\bibitem{Batell:2011pz} 
  B.~Batell, S.~Gori and L.~-T.~Wang,
  ``Exploring the Higgs Portal with 10/fb at the LHC,''
  JHEP {\bf 1206}, 172 (2012)
  \href{http://arxiv.org/abs/1112.5180}{arXiv:1112.5180 [hep-ph]}.
  
\bibitem{Kribs:2012kz} 
  G.~D.~Kribs and A.~Martin,
  ``Enhanced di-Higgs Production through Light Colored Scalars,''
  \href{http://arxiv.org/abs/arXiv:1207.4496}{arXiv:1207.4496 [hep-ph]}.
  
\bibitem{Aad:2011yh} 
  G.~Aad {\it et al.}  [ATLAS Collaboration],
  ``Search for Massive Colored Scalars in Four-Jet Final States in sqrt{s}=7 TeV proton-proton collisions with the ATLAS Detector,''
  Eur.\ Phys.\ J.\ C {\bf 71}, 1828 (2011)
  \href{http://arxiv.org/abs/arXiv:1110.2693}{arXiv:1110.2693 [hep-ex]}.
  
\bibitem{CMS-PAS-EXO-11-016}
 The CMS Collaboration,
  ``Search for New Physics in the Paired Dijet Mass Spectrum,''
   \href{http://cdsweb.cern.ch/record/1416058}{CMS-PAS-EXO-11-016}.  
   
\bibitem{Burgess:2009wm} 
  C.~P.~Burgess, M.~Trott and S.~Zuberi,
  ``Light Octet Scalars, a Heavy Higgs and Minimal Flavour Violation,''
  JHEP {\bf 0909}, 082 (2009)
  \href{http://arxiv.org/abs/0907.2696}{arXiv:0907.2696 [hep-ph]}.
  
\bibitem{Altmannshofer:2012ur} 
  W.~Altmannshofer, R.~Primulando, C.~-T.~Yu and F.~Yu,
  ``New Physics Models of Direct CP Violation in Charm Decays,''
  JHEP {\bf 1204}, 049 (2012)
  \href{http://arxiv.org/abs/arXiv:1202.2866}{arXiv:1202.2866 [hep-ph]}.
  
\bibitem{Carena:2011aa} 
  M.~Carena, S.~Gori, N.~R.~Shah and C.~E.~M.~Wagner,
  ``A 125 GeV SM-like Higgs in the MSSM and the $\gamma \gamma$ rate,''
  JHEP {\bf 1203}, 014 (2012)
  \href{http://arxiv.org/abs/1112.3336}{arXiv:1112.3336 [hep-ph]}.
  
\bibitem{Carena:2012gp} 
  M.~Carena, S.~Gori, N.~R.~Shah, C.~E.~M.~Wagner and L.~-T.~Wang,
  ``Light Stau Phenomenology and the Higgs $\gamma\gamma$ Rate,''
  \href{http://arxiv.org/abs/1205.5842}{arXiv:1205.5842 [hep-ph]}.

\bibitem{Carena:2012xa} 
  M.~Carena, I.~Low and C.~E.~M.~Wagner,
  ``Implications of a Modified Higgs to Diphoton Decay Width,''
  \href{http://arxiv.org/abs/1206.1082}{arXiv:1206.1082 [hep-ph]}.

\bibitem{Joglekar:2012hb} 
  A.~Joglekar, P.~Schwaller and C.~E.~M.~Wagner,
  ``Dark Matter and Enhanced Higgs to Di-photon Rate from Vector-like Leptons,''
  \href{http://arxiv.org/abs/arXiv:1207.4235}{arXiv:1207.4235 [hep-ph]}.
  
\bibitem{An:2012vp} 
  H.~An, T.~Liu and L.~-T.~Wang,
  ``125 GeV Higgs Boson, Enhanced Di-photon Rate, and Gauged U(1)$_{PQ}$-Extended MSSM,''
  \href{http://arxiv.org/abs/1207.2473}{arXiv:1207.2473 [hep-ph]}.
  
\bibitem{Almeida:2012bq} 
  L.~G.~Almeida, E.~Bertuzzo, P.~A.~N.~Machado and R.~Z.~Funchal,
  ``Does $H \to \gamma \gamma$ Taste like vanilla New Physics?,''
  \href{http://arxiv.org/abs/arXiv:1207.5254}{arXiv:1207.5254 [hep-ph]}.
  
\bibitem{Alves:2011kc} 
  A.~Alves, E.~Ramirez Barreto, A.~G.~Dias, C.~A.~de S.Pires, F.~S.~Queiroz and P.~S.~Rodrigues da Silva,
  ``Probing 3-3-1 Models in Diphoton Higgs Boson Decay,''
  Phys.\ Rev.\ D {\bf 84}, 115004 (2011)
  \href{http://arxiv.org/abs/arXiv:1109.0238}{arXiv:1109.0238 [hep-ph]}.
  
\bibitem{Alves:2012yp} 
  A.~Alves, A.~G.~Dias, E.~R.~Barreto, C.~A.~d.~S.~Pires, F.~S.~Queiroz and P.~S.~R.~da Silva,
  ``Explaining the Higgs Decays at the LHC with an Extended Electroweak Model,''
  \href{http://arxiv.org/abs/arXiv:1207.3699}{arXiv:1207.3699 [hep-ph]}.
  
\bibitem{Hall:2011aa} 
  L.~J.~Hall, D.~Pinner and J.~T.~Ruderman,
  ``A Natural SUSY Higgs Near 126 GeV,''
  JHEP {\bf 1204}, 131 (2012)
  \href{http://arxiv.org/abs/arXiv:1112.2703}{arXiv:1112.2703 [hep-ph]}.
  
\bibitem{Blum:2012kn} 
  K.~Blum and R.~T.~D'Agnolo,
  ``2 Higgs or not 2 Higgs,''
  Phys.\ Lett.\ B {\bf 714}, 66 (2012)
  \href{http://arxiv.org/abs/arXiv:1202.2364}{arXiv:1202.2364 [hep-ph]}.
  
\bibitem{Hagiwara:2012mg} 
  K.~Hagiwara, J.~S.~Lee and J.~Nakamura,
  ``Properties of 125 GeV Higgs boson in non-decoupling MSSM scenarios,''
  \href{http://arxiv.org/abs/arXiv:1207.0802}{arXiv:1207.0802 [hep-ph]}.
  
\bibitem{Bellazzini:2012mh} 
  B.~Bellazzini, C.~Petersson and R.~Torre,
  ``Photophilic Higgs from sgoldstino mixing,''
  Phys.\ Rev.\ D {\bf 86}, 033016 (2012)
  \href{http://arxiv.org/abs/arXiv:1207.0803}{arXiv:1207.0803 [hep-ph]}.
  
\bibitem{Craig:2012vn} 
  N.~Craig and S.~Thomas,
  ``Exclusive Signals of an Extended Higgs Sector,''
  \href{http://arxiv.org/abs/arXiv:1207.4835}{arXiv:1207.4835 [hep-ph]}.
  
\bibitem{Alves:2012ez} 
  D.~S.~M.~Alves, P.~J.~Fox and N.~J.~Weiner,
  ``Higgs Signals in a Type I 2HDM or with a Sister Higgs,''
  \href{http://arxiv.org/abs/arXiv:1207.5499}{arXiv:1207.5499 [hep-ph]}.

\end{thebibliography}
\end{document}